\documentclass{aa} 

\usepackage[colorlinks=true, linkcolor=blue, citecolor=blue, urlcolor=blue]{hyperref}

\usepackage[varg]{txfonts}
\usepackage{graphicx}

\usepackage{subcaption}
\usepackage{comment}
\usepackage{booktabs}
\usepackage{bm}
\usepackage{xcolor}
\usepackage{ulem}

\begin{document}

\title{Suppressed diffusion and gamma-ray emission \\ from the Cygnus bubble}
\titlerunning{}
\authorrunning{}

\author{Ben Li\inst{1,2,3} \and Pasquale Blasi\inst{1,2} \and Elena Amato\inst{3}}

\institute{Gran Sasso Science Institute (GSSI), Viale Francesco Crispi 7, 67100 L’Aquila, Italy \and INFN-Laboratori Nazionali del Gran Sasso (LNGS), via G. Acitelli 22, 67100 Assergi (AQ), Italy \and INAF-Osservatorio Astrofisico di Arcetri, Largo E. Fermi 5, 50125 Firenze, Italy}

\date{\today}

\abstract
{Recent gamma-ray observations showed that star clusters (SCs) can be powerful sites of particle acceleration. The spectrum of gamma rays from one such clusters, the Cygnus OB2 association, extends to $\gtrsim$ PeV energies, according to LHAASO measurements. This finding, together with the extended morphology of the emission, leads us to infer that effective particle acceleration up to at least $\gtrsim$1 PeV must be taking place in the SC and that, in the bubble that surrounds it, the diffusivity must be considerably suppressed with respect to typical Galactic values.}
{Here we investigate the gamma-ray emission from the Cygnus region in the case in which particle acceleration takes place at the termination shock of the collective wind of the cluster (WTS) or in an unspecified source at the centre of the bubble, either constant in time or bursting. In the scenario of a collective WTS, particle acceleration and transport can be treated together and the spectrum of accelerated particles can be derived from the transport equation, whereas in the other two cases the accelerator is assumed to be an unspecified source in the SC core. The spectrum and morphology of the gamma-ray emission were computed in all these cases and compared with data by LHAASO.} 
{We numerically solved the transport equation for non-thermal particles in the cases of particle acceleration at the WTS and for a source in the centre of the SC. The spectrum of accelerated particles was computed throughout the bubble so that both the spectrum and the morphology of the gamma-ray emission due to pp scattering could be inferred. We also accounted for the interactions of accelerated particles with the gas outside the bubble, since this contribution may be important to explain the extended region of emission observed by LHAASO. Since penetration of Galactic cosmic rays (GCRs) into the bubble is likely, we also included this phenomenon in the calculations: as a result, some level of shock re-acceleration of GCRs at the WTS is present, for energies $\gtrsim$10 TeV. The predicted gamma-ray emission was then compared with the gamma-ray observations by Fermi-LAT, HAWC and LHAASO.}
{We performed calculations for three different diffusion models and found that a spatially dependent Bohm diffusion coefficient is required in order to account for the spectrum and the morphology of the gamma-ray emission in the case of a collective wind. The GCRs that manage to penetrate the bubble may contribute a significant fraction of the gamma-ray flux above $\sim$ 300 TeV. We also find that a slow diffusion coefficient in a region extending to at least 150 pc from the shock centre is required to reproduce the gamma-ray morphology as measured by LHAASO. The calculations show that in all cases considered here rather extreme assumptions must be made in order to account for the spectrum and morphology as measured by LHAASO.}
{Our study reveals that the diffusion coefficient in the Cygnus bubble must be substantially suppressed with respect to the Galactic average in both cases of particle acceleration at the WTS and for a source in the centre, either constant in time or in the form of a burst. Moreover, the contribution to the ultra-high-energy gamma-ray emission due to GCRs penetrating the cloud around Cygnus is found to be sizable. 
Our conclusion is that explaining both the spectrum and morphology of $\sim$ PeV emission from Cygnus as the result of hadrons accelerated in a non-relativistic steady source requires extreme assumptions.
We also speculate on the possibility that at least some of the highest-energy gamma rays may originate in sources located behind the Cygnus association.}


\maketitle
\nolinenumbers

\section{Introduction}

Star clusters\footnote{Here we will use the word cluster in a loose way, not necessarily indicating a gravitationally bound system.}(SCs) have recently raised the attention of the community, mainly due to the recent detection of ultra-high-energy ($>$100 TeV) gamma rays from the Cygnus OB2 region \cite[]{LHAASO_2024}, with emission extending up to $\sim$ 2 PeV, and the detection of very high-energy ($>$100 GeV) gamma rays from Westerlund 1 by H.E.S.S. \cite[]{Westerlund2022}. Information on the spectrum and morphology of the gamma-ray emission allows us to infer precious information on the origin of the radiation and on the acceleration processes at work in the bubbles produced by these two young SCs. Special interest was stimulated by the LHAASO measurements of the region around Cygnus, in that the gamma-ray spectrum extends to $\gtrsim$ PeV energies, suggesting that particle acceleration up to very high energies must be taking place inside the cluster or in the region around it. Moreover, the emission is observed to spread over a region of $\sim$ 6 degrees around the cluster, corresponding to $\sim$ 150 pc, appreciably larger than the size of the bubble itself.   

This article focuses on the origin of high-energy gamma rays from the Cygnus OB2 bubble and its implications in terms of particle acceleration inside and around the cluster. A wealth of recent work has focused on this problem, in the case of acceleration at the termination shock (TS) of the collective wind of the SC \cite[]{Morlino_2021,Blasi_2023,Menchiari_2023,Menchiari_2024}, and in the cases of a supernova explosion inside the cluster \cite[]{Vieu+2022,Sushch2025}, diffuse second-order acceleration in a static bubble \cite[]{Vieu+2022}, and finally the effect of multiple stellar winds and supernova shocks \cite[]{Voelk_1982,Cesarsky_1983,Webb_1985,Bykov2022}. 

The wind termination shock (WTS) appears to be a very promising acceleration site, in that the upstream of the shock is bound and particle escape cannot take place on that side. However, as discussed by \cite{Morlino_2021}, for reasonable choices of the diffusion coefficient, the spherical symmetry of the problem induces a gradual steepening of the spectrum of accelerated particles, which in fact implies an effective maximum energy that falls short of PeV by at least one order of magnitude. Yet, if particle acceleration at the WTS does occur, the spectrum and morphology of the gamma-ray emission as measured by HAWC \cite[]{Abeysekara_2021} in the direction of Cygnus OB2 is very well reproduced in terms of hadronic interactions with gas in the bubble \cite[]{Blasi_2023}. The recent results of LHAASO \cite[]{LHAASO_2024} show that the gamma-ray spectrum and the emission region extend, respectively, to the PeV range and to a radius much larger than that of the bubble itself, challenging the scenario in which particle acceleration occurs at the WTS alone. The problem is made even more complex by the results of recent hydrodynamical simulations of the bubble around this SC, suggesting that the formation of the WTS is not guaranteed \cite[]{Vieu_2024} due to the rather broad spatial distribution of the stars in the cluster. As a result of all these considerations, alternative scenarios for the origin of the accelerated particles and their associated gamma-ray emission should be considered. 

While the possibility that magnetic turbulence in the bubble may energize particles through stochastic acceleration is interesting from the phenomenological point of view \cite[]{Vieu+2022}, this mechanism cannot account for the acceleration of particles to $E\gtrsim$ PeV. Moreover, whether this process does or does not have implications depends on the details of wave propagation. Hence, in the following, we do not consider this case in any greater detail. 

The three cases that we focus on are the following: 1) particle acceleration at a putative TS of the collective wind, if one is formed; 2) particle acceleration at a source in the centre of the SC, assumed to be stationary on the timescales on which the cluster evolves (approximately millions of years). This might be the case for winds of individual stars or collisions between winds of individual stars; and 3) a bursting source in the centre of the cluster, as might be the case for a supernova that exploded some time ago in the cluster core. Of these three cases, the only one in which the maximum energy of accelerated particles can be calculated self-consistently within our set of assumptions, once the physical properties of turbulence in the bubble are assumed, is that of the TS, where acceleration is part of the transport of particles in the collective wind. In the other two cases, we can only use plausibility arguments to estimate the maximum energy. For instance, the case of a SN explosion in a cluster was considered by \cite{Harer_2025}, but the maximum energy was estimated based on a simple application of the Hillas criterion. 

This paper is organized as follows. In Section \ref{sec:max energy} we discuss particle acceleration at the WTS of the collective wind, if one is formed, and discuss the gamma-ray spectrum and morphology to be expected. The penetration of Galactic cosmic rays (GCRs) into the region and its implications in terms of gamma-ray emission are discussed. In Section \ref{sec:cont}, we discuss the case of a continuous source at the centre of the cluster, while the case of a burst-like source is discussed in Section \ref{sec:burst}. In all these cases the role of penetrating GCRs is discussed in detail. Our conclusions in terms of suppressed diffusivity and reliability of the acceleration processes are presented in Section \ref{sec:discuss}.

\section{Particle acceleration at the WTS}
\label{sec:max energy}

If the SC is sufficiently compact at a given time, the winds of the individual stars can form a collective wind that plows the surrounding material outwards and forms a rarefied hot bubble. While the pre-existing gas is expected to accumulate at the edge of this bubble, gas cooling and fragmentation can be presumed to leave behind clouds of dense molecular and atomic gas, co-spatial with the hot rarefied collective wind. If this is the case, the wind excavates a bubble with radius \cite[]{Dyson_1972, Castor_1975, Weaver_1977, Gupta_2018,Morlino_2021} (see Appendix \ref{app:evolution} for details):
\begin{equation}
    R_{\mathrm{b}}(t) \simeq 139 ~ n_{10}^{-1/5} ~ \dot{M}_{-4}^{1/5} ~ v_8^{2/5} ~ t_{10}^{3/5} ~ \mathrm{pc},
    \label{eq:bubble radius}
\end{equation}
where $n_{10}=n_0/(10 ~ \mathrm{cm}^{-3})$ is the number density of the ambient gas, $\dot{M}_{-4}=\dot{M}/(10^{-4} ~ M_{\odot} ~ \mathrm{yr}^{-1})$ is the total mass loss rate of a SC, and $v_8=v_\mathrm{w}/(1000 ~ \mathrm{km} ~ \mathrm{s}^{-1})$ is the uniform wind velocity upstream of the shock. The quantity $t_{10}=T_\mathrm{age}/(10 ~ \mathrm{Myr})$ is  the dynamical age of the system. For Cygnus OB2, we use $n_0 = 20 ~ \mathrm{cm}^{-3}$, $\dot{M} = 1.5 \times 10^{-4} ~ M_{\odot} ~ \mathrm{yr}^{-1}$, $v_\mathrm{w} = 2800 ~ \mathrm{km} ~ \mathrm{s}^{-1}$, and $T_\mathrm{age} = 3 ~ \mathrm{Myr}$ throughout this paper, which result in $R_{\mathrm{b}} \sim 96 ~ \mathrm{pc}$.
We note that these are typical parameter values for Cygnus OB2, although some level of uncertainty may exist. For example, recent studies indicate that the ambient gas number density is of the order of $\sim 10~\mathrm{cm}^{-3}$; however, it is still subject to large uncertainties due to the complexity of the environment \citep{Astiasarain_2023,Menchiari_2023,Menchiari_2024,Harer_2025}.

Since the wind is supersonic, its bulk energy must be dissipated at a TS, whose radius can be estimated by equating the ram pressure of the upstream wind with the pressure inside the bubble, yielding
\begin{equation}
    R_{\mathrm{s}}(t) \simeq 24.3 ~ n_{10}^{-3/10} ~ \dot{M}_{-4}^{3/10} ~ v_8^{1/10} ~ t_{10}^{2/5} ~ \mathrm{pc},
    \label{eq:shock radius}
\end{equation}
which corresponds to $R_{\mathrm{s}} \sim 15 ~ \mathrm{pc}$ for Cygnus OB2. In this picture, particle acceleration occurs at the TS and the downstream medium is advected away from the shock with a radial dependent velocity, $u(r)\propto r^{-2}$. 

If one assumes that a fraction, $\eta_B<1$, of the wind kinetic energy is converted into turbulent magnetic energy, such that $B^2/4\pi=\eta_B \rho u^2$, one can estimate the magnetic field immediately upstream of the TS as
\begin{equation}
    B_1(R_\mathrm{s}) \simeq 10.6 ~ \eta_B^{1/2} ~ n_{10}^{3/10} ~ \dot{M}_{-4}^{1/5} ~ v_8^{2/5} ~ t_{10}^{-2/5} ~ \mu \mathrm{G}.
    \label{eq:magnetic field}
\end{equation}
Here we assume that the magnetic field is fully turbulent, with no mean field component. Immediately downstream of the shock we assume that the magnetic field strength, $B_2(R_\mathrm{s})$, is enhanced by a factor of $\sqrt{11}$, appropriate for a strong shock \cite[]{Morlino_2021}.
For reasonable values of $\eta_B$, one can see that typical magnetic field strength at the shock is $\lesssim$ few $\mu$G. Unless otherwise specified, $\eta_B$ is set to 0.1 throughout the paper, because as we discuss below, lower values do not lead to a satisfactory description of the data. However, it is worth stressing that this is an extreme assumption, in that $\eta_B=0.1$ corresponds to Alfv\'enic Mach number of $\sim 3$, in which case shock acceleration may not be very efficient.

The diffusion coefficient in the bubble can be estimated by assuming a Kolmogorov or a Kraichnan phenomenology, in which cases: 
\begin{align}
    D_{\mathrm{Kol}}(E) = \frac{1}{3} c L_{\mathrm{c}}^{2/3} r_{\mathrm{L}}^{1/3}, \\
    D_{\mathrm{Kra}}(E) = \frac{1}{3} c L_{\mathrm{c}}^{1/2} r_{\mathrm{L}}^{1/2}, 
\end{align}
respectively, with $L_{\mathrm{c}}$ the coherence scale of the perturbations, which is assumed here to be of the same order of magnitude as the average separation between massive stars in the cluster, $L_{\mathrm{c}} \simeq \langle d \rangle \simeq 1$ pc. While slightly larger values would still be plausible, they would imply a larger diffusion coefficient, and hence a lower maximum energy of accelerated particles. As we discuss below, even assuming optimistic values for these parameters, providing a good description of observations is still challenging.
We stress that the expressions above are to be used only for energies such that $r_\mathrm{L}<L_\mathrm{c}$, while at higher energies the mean free path scales $\lambda \propto E^2$, which implies that $D(E)\propto E^2$ if no larger-scale structure exists aside from the assumed turbulent magnetic field.
If the prescription $B^2\propto \rho u^2$ is applied to the whole downstream region (see below), then $B(r)\propto r^{-2}$, and for a given energy the Larmor radius increases as $r_\mathrm{L}\propto r^2$.
In this case, the transition from the inertial range to the quasi-ballistic regime for a particle of given energy $E$ occurs at a distance downstream, $R_\mathrm{t}(E) = R_\mathrm{s} [e B_2(R_\mathrm{s}) L_\mathrm{c} / E]^{1/2}$. For a particle of energy $E=1~\mathrm{PeV}$, we have $R_\mathrm{t}\sim 85~\mathrm{pc}$ for the parameter values adopted above. The effect of the transition is negligible, since $R_\mathrm{t}(E_\mathrm{max})$ is comparable to $R_\mathrm{b}$. Therefore, for simplicity, we assume that particles always diffuse in the $r_\mathrm{L}<L_\mathrm{c}$ regime.

Below we also consider the rather extreme case of Bohm diffusion, $D_{\mathrm{Bohm}} = (1/3) c r_{\mathrm{L}}$, keeping in mind, however, that it is physically meaningful only in those situations in which the turbulence is self-generated and the spectrum of the accelerated particles is the canonical $\propto E^{-2}$ (for relativistic particles). For SCs it is not guaranteed that this is the case (see discussion in \cite{Morlino_2021,Blasi_2023}).

Whether these diffusion coefficients are spatially constant or not depends on whether one assumes that turbulence is produced at some location and advected or rather generated at each location. In either case this is simply an assumption, since we are following neither the cascading of turbulence in $k$ space here nor its advection. Below we numerically solve the transport equation for non-thermal particles with the same technique previously discussed by \cite{Blasi_2023}:
\begin{equation}
    \frac{1}{r^2} \frac{\partial}{\partial r} \left[r^2 D \frac{\partial f}{\partial r}\right] - \frac{1}{r^2} \frac{\partial}{\partial r} \left[r^2\tilde{u} f\right] + \frac{1}{p^2} \frac{\partial}{\partial p} \left[p^2\dot{p} f\right] + Q = \frac{\partial f}{\partial t},
\end{equation}
where $f(r,p,t)$ is the particle distribution function in phase space, for the two cases of spatially uniform and space dependent diffusion. The function $Q(r,p,t)$ represents the injection term, which can be expressed by $Q(r,p,t)=4 \pi R_{\mathrm{s}}^2 \eta_{\mathrm{inj}} n_1 u_1 [\delta(p - p_{\mathrm{inj}})/(4 \pi p^2)] [\delta(r - R_{\mathrm{s}})/(4 \pi r^2)]$. The energy loss term includes both the adiabatic energy loss/gain and the energy loss by interactions; namely, $\dot{p}=\dot{p}_{\mathrm{ad}}+\dot{p}_{\mathrm{int}}$. $D(r,p)$ and $\tilde{u}(r)=u+\eta v_{\mathrm{A}}$ are the diffusion coefficient and the effective advection velocity, respectively. The coefficient $\eta$ mimics the effect of a net motion of the scattering centres; namely, the possibility of a non-vanishing local Alfv\'en speed \cite[]{Blasi_2023}. Here we assume that waves propagate isotropically in the upstream ($\eta_1=0$), while anisotropy can exist in the downstream ($\eta_2\neq0$). The transport equation was solved numerically, with a zero flux boundary condition at $r=0$ and free escape boundary at $r=R_\mathrm{b}$ (see Appendix \ref{app:transport} for details). As is discussed below, the latter boundary condition must be changed if one intends to include the penetration of GCRs into the bubble. We assume that particle transport has reached steady-state, which is justified for most of the cases in our calculations.

\subsection{Uniform and non-uniform diffusion in the superbubble}

The possibility of accounting for a diffusion coefficient that is not spatially uniform downstream of the TS was implemented here by assuming that the energy density of the turbulent spectrum, $\delta B^2/4 \pi$, is a fixed fraction of the local ram pressure, $\rho u^2$. Since downstream $u_2(r)\propto r^{-2}$, it follows that $\delta B^2\propto r^{-4}$, which reflects in the diffusion coefficient increasing as $D_2(r,E)\propto r^{2\delta}$, where $\delta=1/3,~1/2,~1$ for Kolmogorov, Kraichnan, and Bohm diffusion, respectively. In the homogeneous case, on the other hand, we assume that the diffusion coefficient downstream equals its value immediately behind the TS (see \cite{Blasi_2023}).
The spatial dependence of $D_2$ allows for the easier escape of particles from downstream, compared to the case of homogeneous diffusion, so that the gamma-ray morphology is also affected.
On the other hand, we neglect the spatial dependence of the diffusion coefficient in the upstream region, and use its value immediately upstream of the shock. This is a good approximation, since $D_1(r,E) \propto r^\delta$ has a weaker dependence on $r$ than $D_2$.

The inhomogeneity of the diffusion coefficient has implications for both the maximum energy of the accelerated particles and for the morphology and spectrum of the gamma-ray emission from the superbubble. However, it can be understood rather easily that while inhomogeneous diffusion leads to a somewhat lower maximum energy, the effect is not prominent, since the acceleration process occurs close to the TS. The analytic solution of particle transport in the bubble as derived by \cite{Morlino_2021} provides us with a direct estimate of the maximum energy in the form of an implicit condition:
\begin{equation}
    \int_{R_\mathrm{s}}^{R_\mathrm{b}} \frac{u_2(r)}{D_2(r,E_{\mathrm{max}})} dr \simeq 1,
    \label{eq:max energy}
\end{equation}
where $D_2(r,E)$ is the diffusion coefficient in the downstream region. When $D_2$ is spatially constant, this reduces to
\begin{equation}
    \frac{u_2 R_{\mathrm{s}}}{D_2(E_{\mathrm{max}})} \left(1-\frac{R_\mathrm{s}}{R_\mathrm{b}}\right) \simeq 1,
\end{equation}
and for $R_\mathrm{s}$ appreciably smaller than $R_\mathrm{b}$ the condition is equivalent to assuming that the size of the downstream diffusion region becomes comparable to the size of the TS. For inhomogeneous diffusion, Equation \eqref{eq:max energy} results in 
\begin{equation}
    \frac{1}{2\delta +1} \frac{u_2 R_{\mathrm{s}}}{D_2(E_{\mathrm{max}})} \left[1-\left(\frac{R_\mathrm{s}}{R_\mathrm{b}}\right)^{2\delta +1}\right] \simeq 1,
\end{equation}
and again for $R_\mathrm{s}<R_\mathrm{b}$, it leads to basically the same condition, modified by a numerical factor of order unity. In conclusion, for the same choice of the diffusion coefficient, the introduction of a spatial dependence downstream changes the maximum energy by a relatively small factor. On the other hand, we expect the spectrum and morphology of the gamma-ray emission to be severely affected by inhomogeneous diffusion.  

Accelerated protons propagate in the bubble because of advection and diffusion, and interact with the target gas, producing hadronic gamma-ray emission. The total gamma-ray flux from a cluster was calculated by integrating the gamma-ray emissivity over the volume of the bubble, as detailed in Appendix \ref{app:emission}. Here, we adopted $n_\mathrm{b}=10 ~ \mathrm{cm}^{-3}$ for the total gas number density inside the bubble, as contributed by atomic and molecular gas left inside as a consequence of cooling or as remnants of dense clouds after the passage of the shock. 

\begin{figure}[t]
    \centering
    \begin{subfigure}[b]{0.48\textwidth} 
    \centering
    \includegraphics[width=\textwidth]{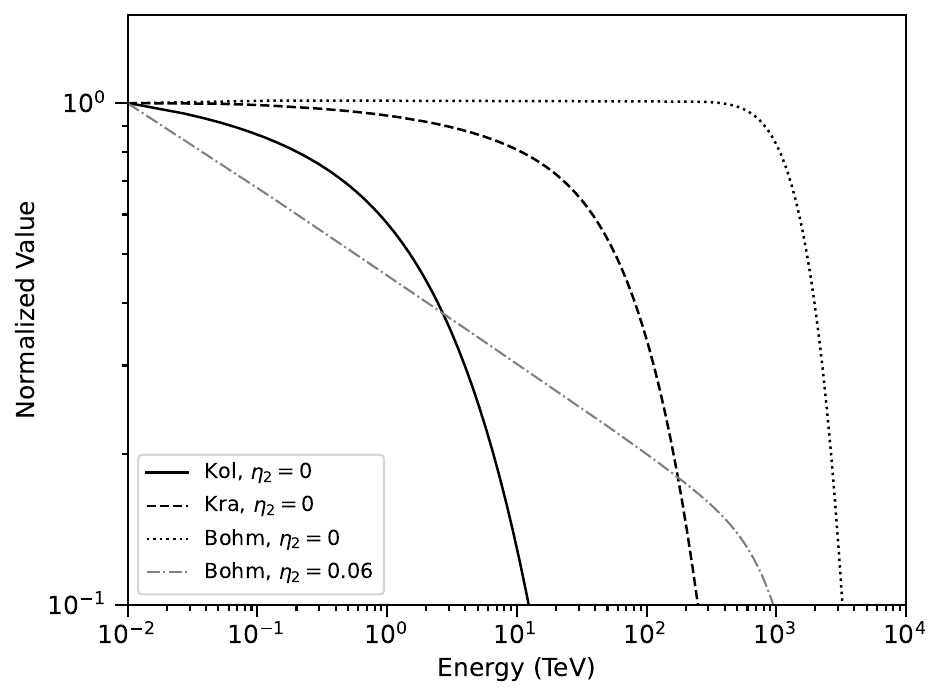} 
    \end{subfigure}
    \hfill 
    \begin{subfigure}[b]{0.48\textwidth}
    \centering
    \includegraphics[width=\textwidth]{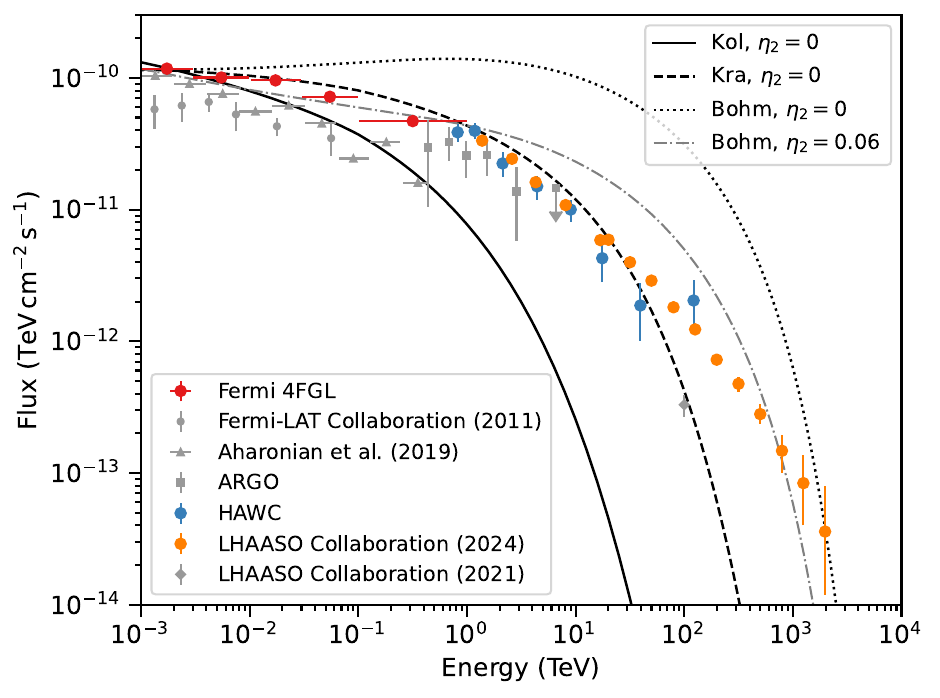} 
    \end{subfigure}
    \caption{Particle spectrum at the shock (top) and total gamma-ray emission from the Cygnus bubble (bottom), for uniform Kolmogorov, Kraichnan, and Bohm diffusion coefficients ($\eta_2=0,~0.06$), respectively. The red dots represent the latest GeV data from Fermi-LAT, while earlier data are shown as grey dots and grey triangles \citep{Abdollahi_2020,Ackermann_2011,Aharonian_2019}. The grey squares refer to ARGO \citep{Bartoli_2014}. The blue dots and orange dots represent the TeV gamma-ray measurements by HAWC and LHAASO \citep{Abeysekara_2021,LHAASO_2024}. Earlier LHAASO data are shown as grey diamonds \citep{Cao_2021}.}
    \label{fig:uniform diff}
\end{figure}

\begin{figure}[t]
    \centering
    \begin{subfigure}[b]{0.48\textwidth} 
    \centering
    \includegraphics[width=\textwidth]{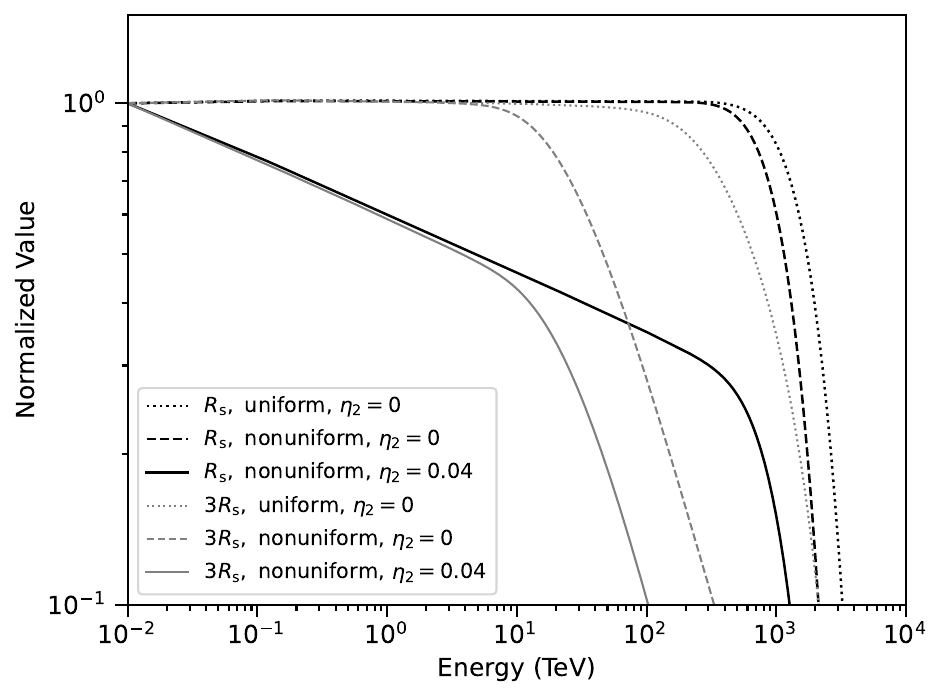} 
    \end{subfigure}
    \hfill 
    \begin{subfigure}[b]{0.48\textwidth}
    \centering
    \includegraphics[width=\textwidth]{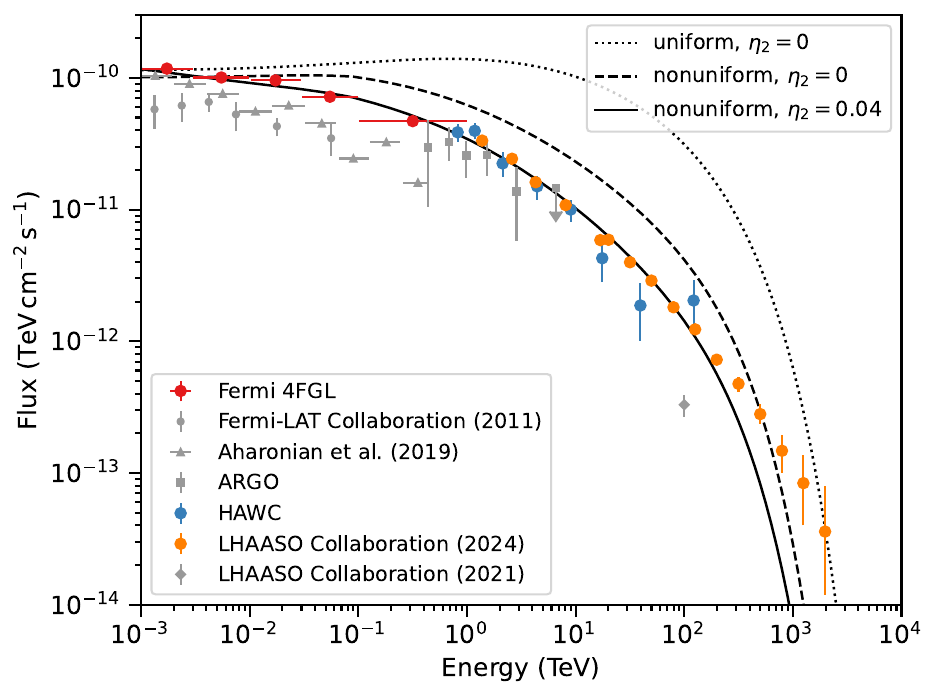} 
    \end{subfigure}
    \caption{Particle spectrum at $R_{\mathrm{s}}$ and $3R_{\mathrm{s}}$ (top), and total gamma-ray emission from the Cygnus bubble (bottom), for uniform Bohm diffusion (dotted lines), non-uniform Bohm diffusion with $\eta_2=0$ (dashed lines), and non-uniform Bohm diffusion with $\eta_2=0.04$ (solid lines).}
    \label{fig:nonuniform diff}
\end{figure}

The top panel of Figure \ref{fig:uniform diff} shows the calculated particle spectrum at the WTS of Cygnus OB2 for the uniform Kolmogorov, Kraichnan, and Bohm diffusion coefficients in the bubble. One can easily see that the spectrum of accelerated particles at the shock drops smoothly towards the maximum energy in the cases of Kolmogorov and Kraichnan diffusion, while the flux suppression at the maximum energy is sharper in the case of Bohm diffusion. This effect has already been discussed by \cite{Morlino_2021} as being due to the spherical symmetry of the problem. 

The bottom panel of Figure \ref{fig:uniform diff} shows the predicted total gamma-ray flux from the Cygnus bubble, overlapping with the gamma-ray data points measured by Fermi-LAT, ARGO, HAWC, and LHAASO. For the Kolmogorov and Kraichnan diffusion coefficients, the maximum energy is too low to explain the gamma-ray data. Although the maximum energy is large enough in the case of Bohm diffusion, the shape of the gamma-ray spectrum is quite unlike the observed one. The slope of the spectrum is affected by $\eta_2$ through the effective compression factor, $\mathcal{R}$ (see Appendix \ref{app:transport}); however, $\eta_2$ does not affect the global shape of the gamma-ray emission. 

This figure shows that while homogeneous diffusion in the bubble may properly describe the gamma-ray spectrum as observed by Fermi-LAT, ARGO, and HAWC for the case of Kraichnan diffusion, it cannot account for its highest-energy data point, at $E\gtrsim$20 TeV. On the other hand, Bohm diffusion does not properly describe the spectral shape of the gamma-ray emission. 

In the case of inhomogeneous diffusion, it becomes even more evident that the maximum energy of accelerated particles for Kolmogorov and Kraichnan diffusion is too low to account for observations; hence, in that case we focus on Bohm diffusion. However, we reiterate that this scenario is difficult to justify in the absence of efficient self-generation, which does not appear to be well justified in superbubbles (see discussion by \cite{Blasi_2023}). Our results for the case of inhomogeneous diffusion are illustrated in Figure \ref{fig:nonuniform diff}: the top panel shows the particle spectra at the shock ($r=R_{\mathrm{s}}$) and at a larger radius ($r=3R_{\mathrm{s}}$), for three different Bohm diffusion scenarios. For Bohm diffusion with $\eta_2=0$, the maximum energy at the shock is slightly lower than that in the uniform Bohm diffusion case, as expected from the discussion above. However, the spectrum drops rapidly with increasing radius for non-uniform Bohm diffusion. We also plot the spectrum for Bohm diffusion and $\eta_2=0.04$. As can be seen, $\eta_2$ only affects the slope of the spectrum.

The bottom panel of Figure \ref{fig:nonuniform diff} shows the comparison between the predicted gamma-ray spectrum for uniform and non-uniform Bohm diffusion in the downstream. For non-uniform Bohm diffusion with $\eta_2=0.04$, our predicted gamma-ray emission fits the gamma-ray data well below $\sim$ 300 TeV, but it is still hard to account for the $\sim$ PeV photons measured by LHAASO. This may indicate the presence of other acceleration mechanisms, or gamma-ray contributions from other sources, either inside the cluster or behind it (such as Cygnus X-3 \citep{Kachelriess_2025, LHAASO_cygx3}), or a residual contribution due to the GCR background, as discussed below.

\subsection{Contribution of GCRs to gamma-ray emission in the Cygnus bubble}
\label{sec:penetration}

\begin{figure}[t]
    \centering
    \includegraphics[width=0.48\textwidth]{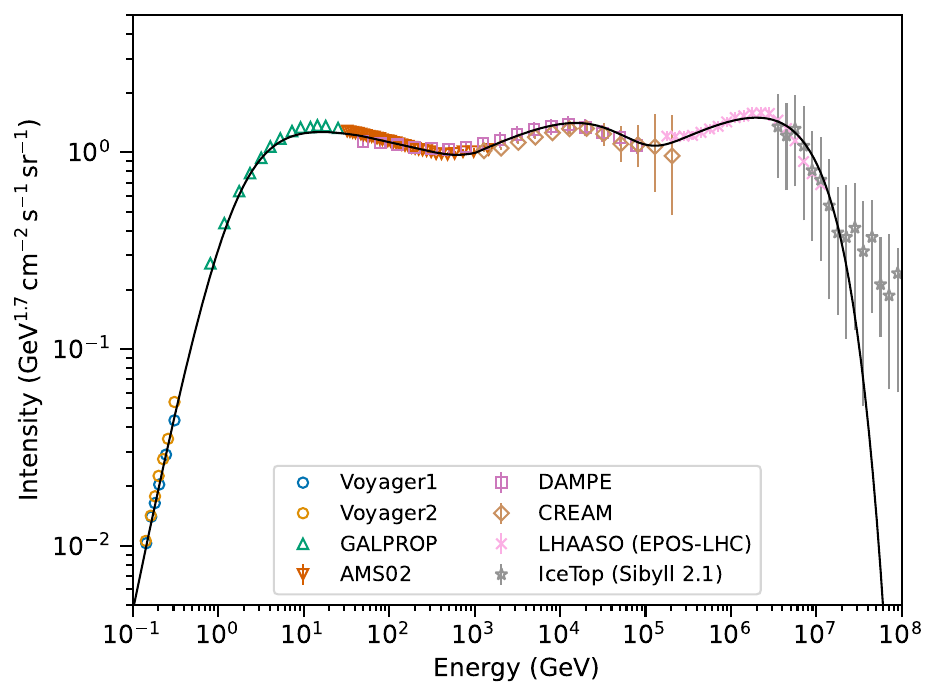} 
    \caption{Galactic cosmic-ray proton spectrum as measured in the local ISM. The black line represents a fit based on the parameterizations in \cite{Shen_2021,Lipari_2020,Cao_2023}. The blue and orange circles are the proton spectra measured by Voyager 1 and Voyager 2, respectively \citep{Cummings_2016, Stone_2019}. The green triangles are the proton flux computed with GALPROP \citep{Bisschoff_2019}. The red triangles, magenta squares, and brown diamonds represent the measurements of AMS-02, DAMPE, and CREAM \citep{Aguilar_2015,An_2019,Yoon_2017}. The pink crosses are the proton flux measured by LHAASO with the hadronic interaction model of EPOS-LHC \citep{LHAASO_2025}. The grey stars represent the measurements of IceTop with the hadronic interaction model of Sibyll 2.1 \citep{Aartsen_2019}.}
    \label{fig:gcr}
\end{figure}

\begin{figure*}[t]
    \centering
    \begin{subfigure}[b]{0.48\textwidth} 
    \centering
    \includegraphics[width=\textwidth]{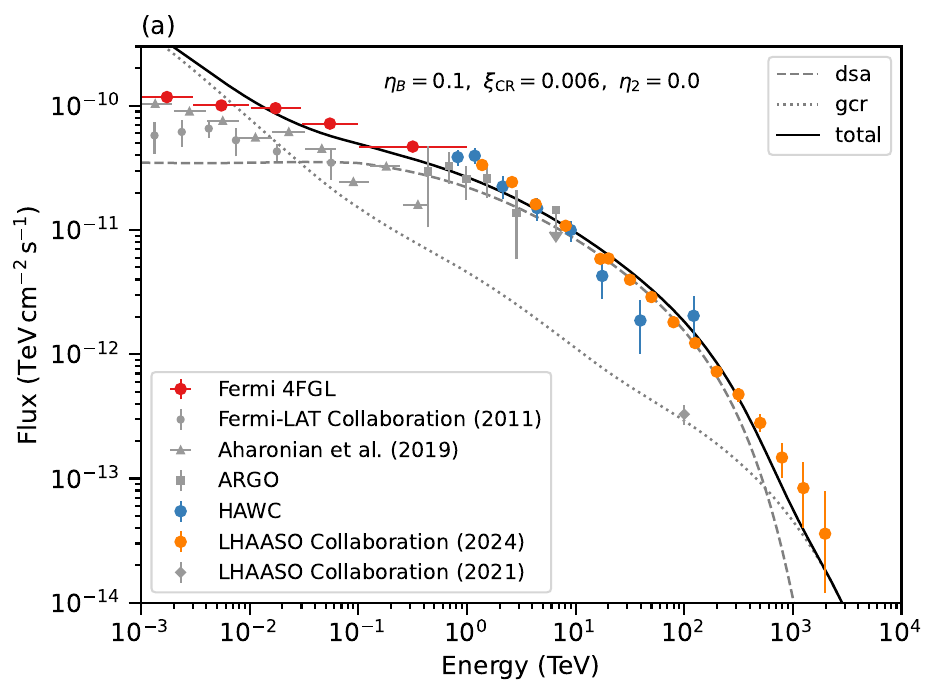} 
    \end{subfigure}
    \hfill 
    \begin{subfigure}[b]{0.48\textwidth} 
    \centering
    \includegraphics[width=\textwidth]{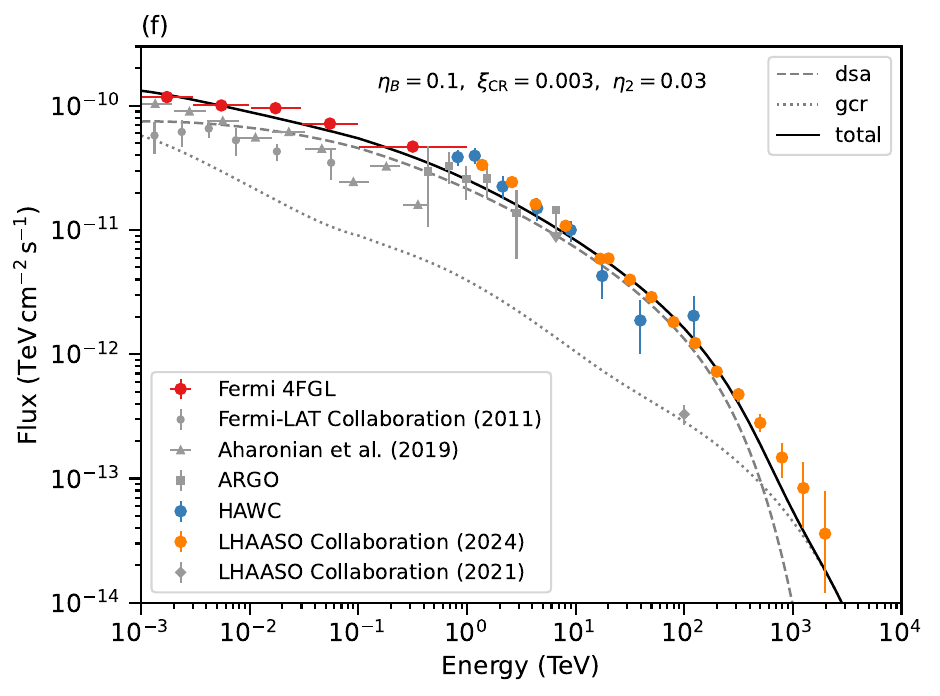} 
    \end{subfigure}
    
    \begin{subfigure}[b]{0.48\textwidth}
    \centering
    \includegraphics[width=\textwidth]{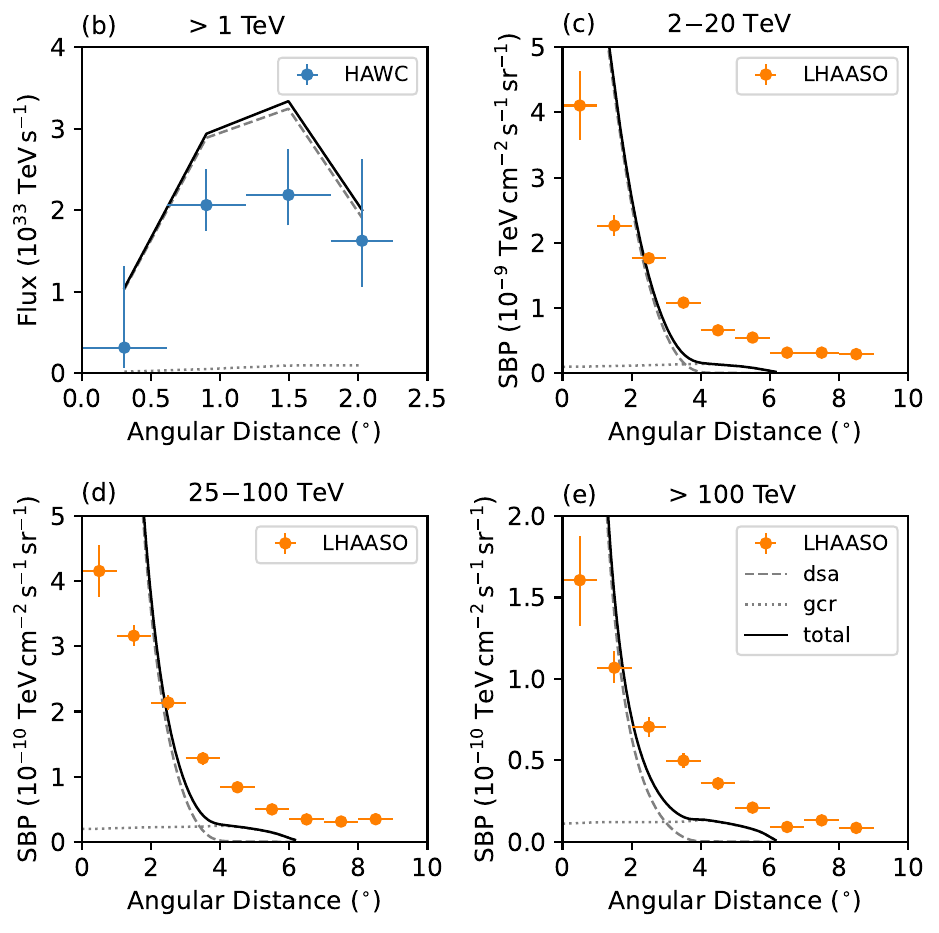} 
    \end{subfigure}
    \hfill 
    \begin{subfigure}[b]{0.48\textwidth}
    \centering
    \includegraphics[width=\textwidth]{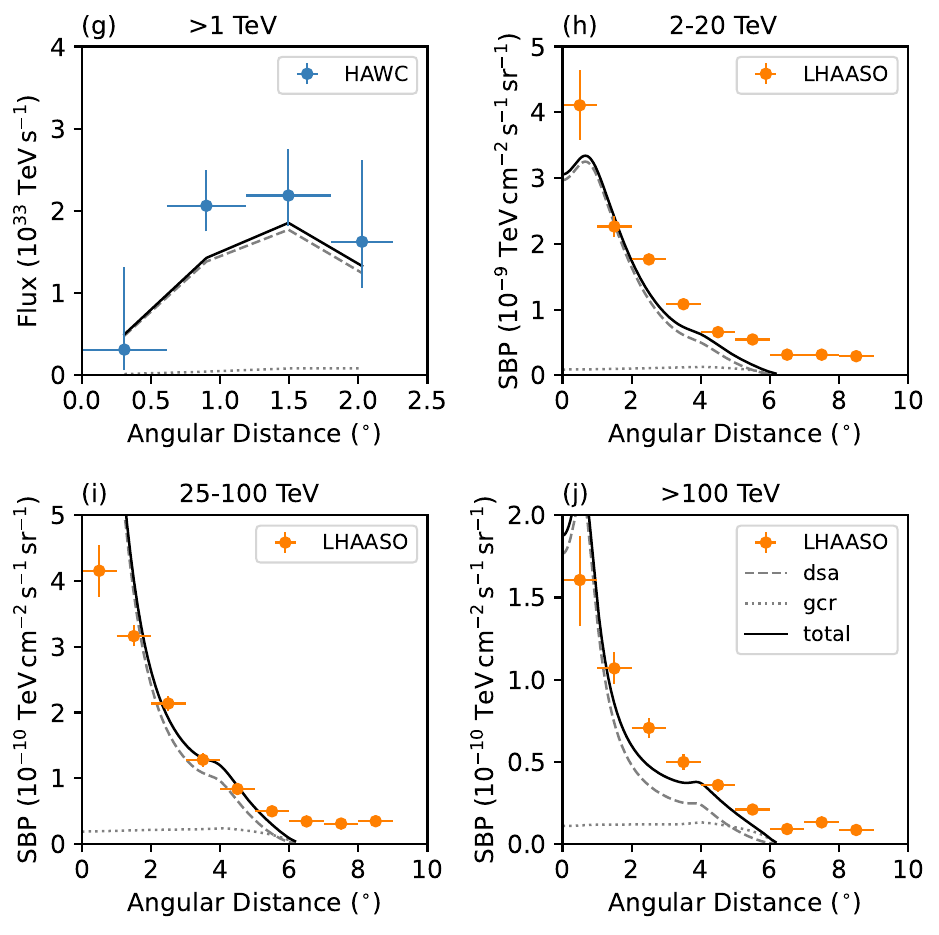} 
    \end{subfigure}
    \caption{Total gamma-ray flux from the Cygnus bubble (top) and gamma-ray morphologies in different energy ranges (bottom), for two cases: (1) Bohm diffusion in the bubble and Galactic diffusion in the cloud (left panels), and (2) Bohm diffusion in the bubble and slow diffusion in the cloud with $D_{\mathrm{c}}(E) = 4 \times 10^{24} ~ E_{\mathrm{GeV}}^{0.6} ~ \mathrm{cm}^{2} ~ \mathrm{s}^{-1}$ (right panels). The dashed lines represent the gamma-ray emission from shock accelerated particles, while the dotted lines represent the emission from penetrated GCRs.}
    \label{fig:fast & slow diffusion}
\end{figure*}

\begin{figure}[t]
    \centering
    \begin{subfigure}[b]{0.48\textwidth} 
    \centering
    \includegraphics[width=\textwidth]{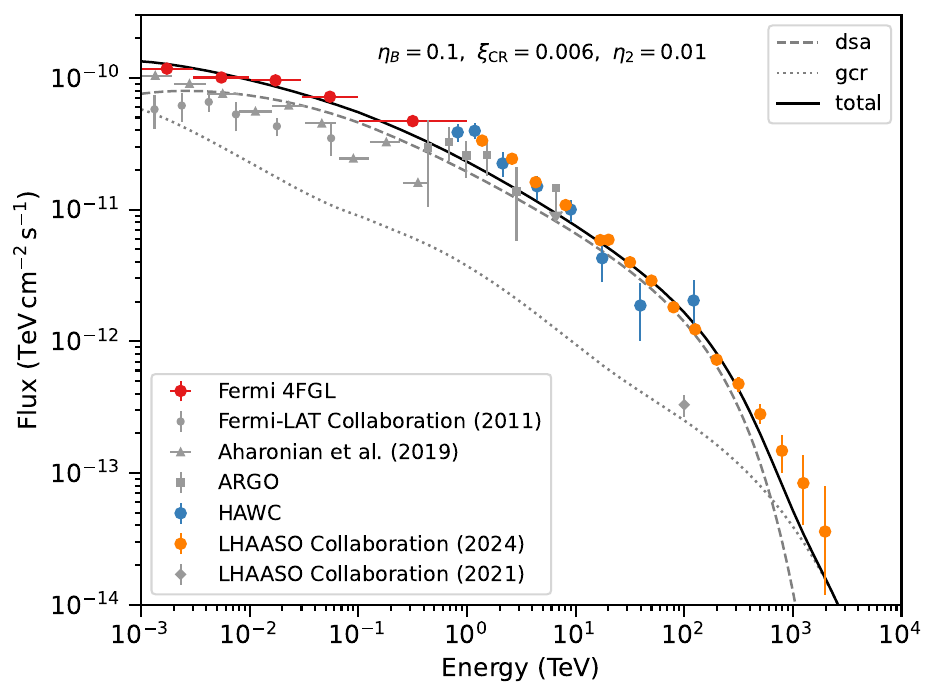} 
    \end{subfigure}
    
    \begin{subfigure}[b]{0.48\textwidth}
    \centering
    \includegraphics[width=\textwidth]{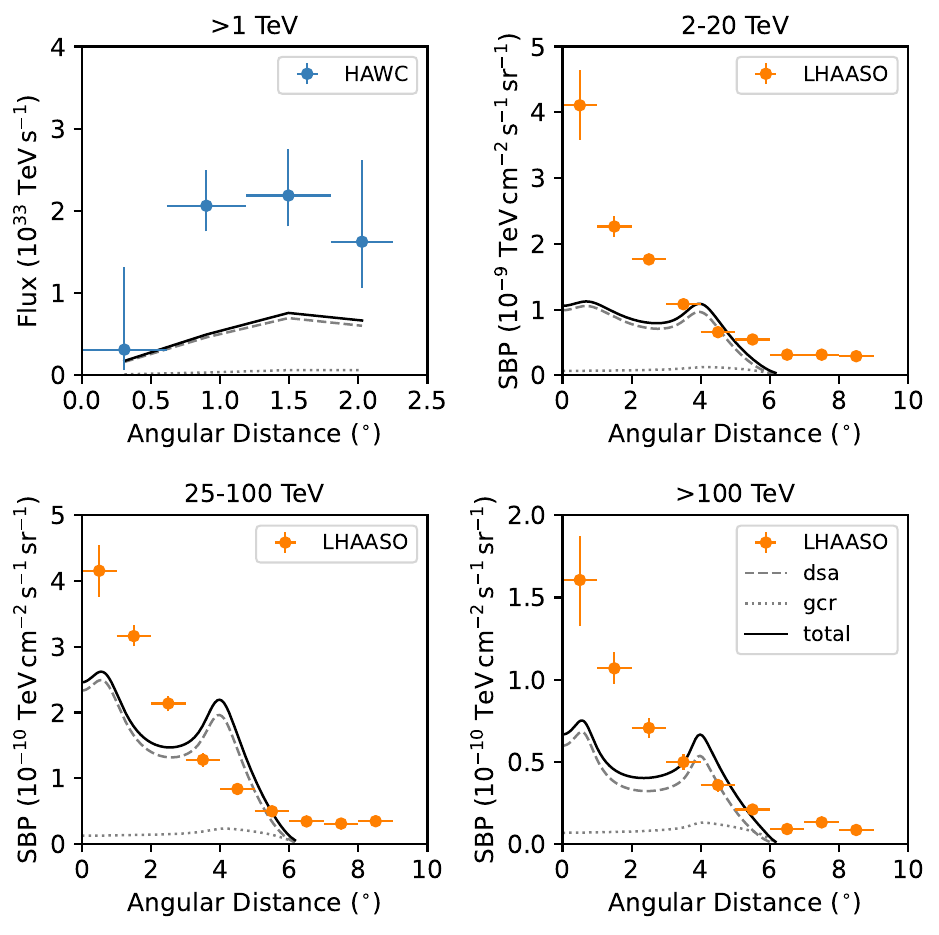} 
    \end{subfigure}
    \caption{As in Figure \ref{fig:fast & slow diffusion} for slow diffusion in the cloud, but assuming a lower bubble gas density of $n_{\rm b}=1 ~ \rm{cm}^{-3}$.}
    \label{fig:low density bubble}
\end{figure}

\begin{figure}[t]
    \centering
    \begin{subfigure}[b]{0.48\textwidth} 
    \centering
    \includegraphics[width=\textwidth]{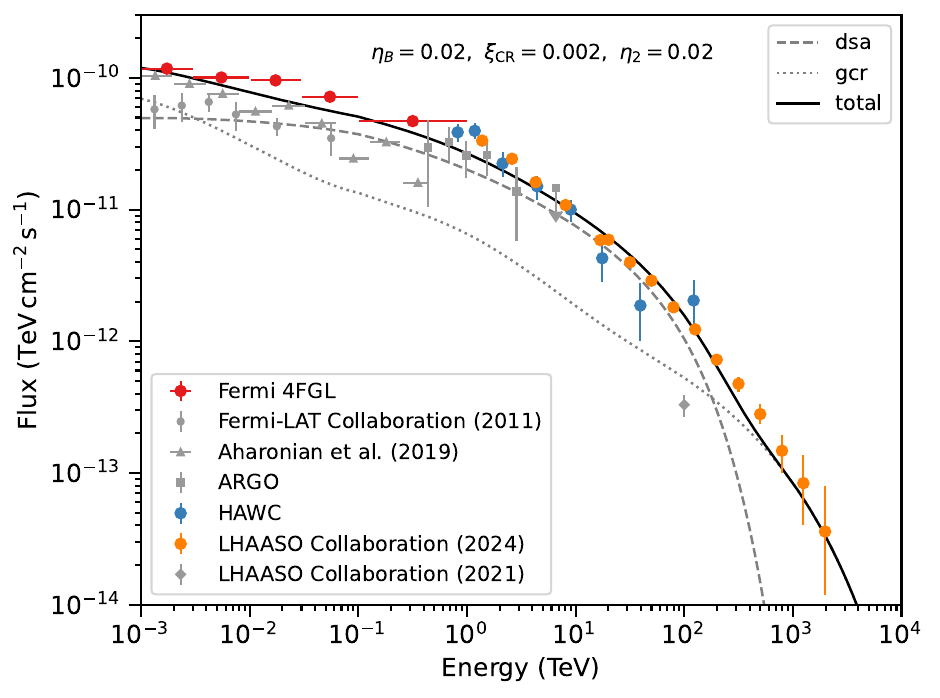} 
    \end{subfigure}
    
    \begin{subfigure}[b]{0.48\textwidth}
    \centering
    \includegraphics[width=\textwidth]{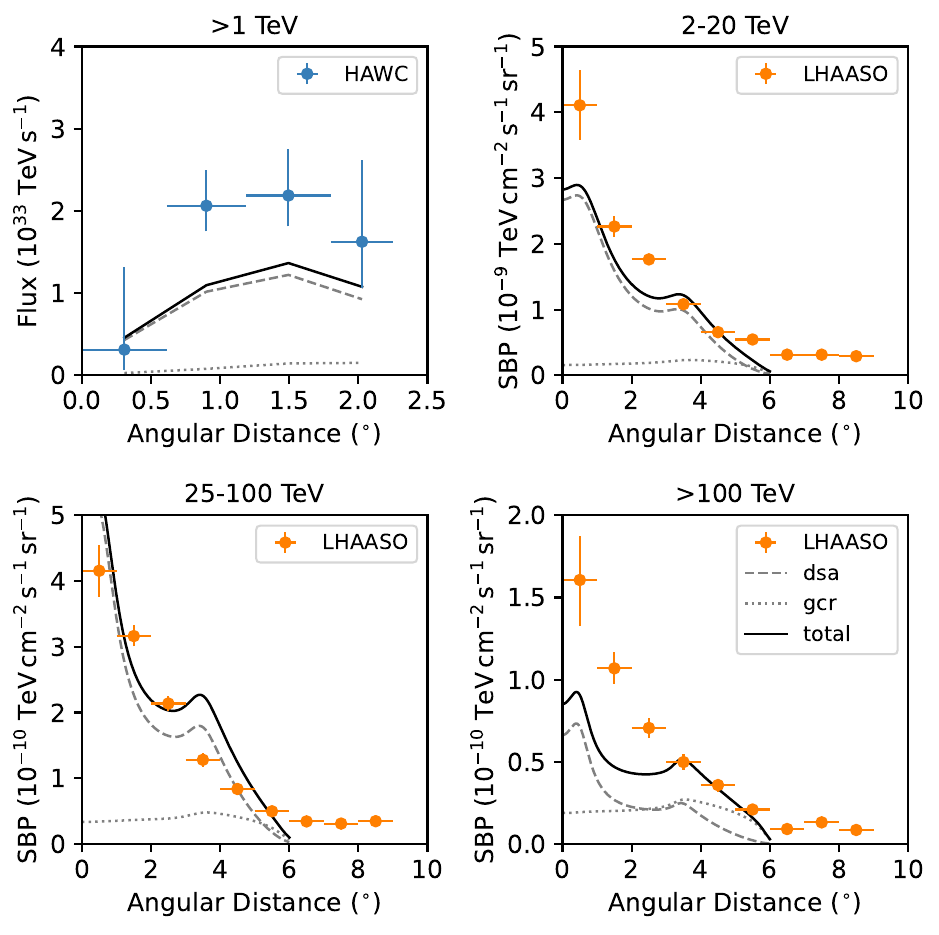} 
    \end{subfigure}
    \caption{As in Figure \ref{fig:fast & slow diffusion} for slow diffusion in the cloud, but using a lower magnetic efficiency of $\eta_B=0.02$ and a higher gas density in the cloud $n_{\rm c}=40 ~ \rm{cm}^{-3}$.}
    \label{fig:lower etaB}
\end{figure}

Galactic cosmic rays can contribute to the gamma-ray emission of the Cygnus region in three ways: 1) penetration inside the bubble and interaction with local gas; 2) interactions of GCRs with the gas outside the bubble; and 3) penetration and re-energization of GCRs, for those particles that manage to reach the TS, against the advective flow.

In order to assess the role of GCRs in the Cygnus region, it is useful to introduce the morphology of the gamma-ray emission: recently, LHAASO has measured the gamma-ray intensity profile of the Cygnus bubble in three different energy ranges: 2--20 TeV, 25--100 TeV, and >100 TeV \citep[see][]{LHAASO_2024}. These measurements indicate that the emission extends to at least 6 degrees from the bubble centre, corresponding to a radius of about 150 pc at a distance of 1.4 kpc. The gamma-ray intensity remains nearly constant at angular radii greater than 6 degrees for all three energy ranges, suggesting that perhaps the emission is caused by a homogeneous distribution of CR particles. From Equation \eqref{eq:bubble radius}, the radius of the Cygnus bubble may be computed as $\sim$ 96 pc for typical parameter values. This radius, which is likely larger than the actual bubble since our estimate neglects the effects of cooling, is significantly smaller than that inferred from LHAASO observations.

The most straightforward implication of this piece of observation is that there is target gas in the region outside the bubble, perhaps associated with the gas plowed away by the forward shock, although the thickness of such region seems too large to be interpreted in that way. This gas could also be part of the cloud from which the cluster formed several million years ago. Below we assume that there is a cloud of gas that extends to $\sim 150~\mathrm{pc}$ from the centre with a density of $n_\mathrm{c} \sim n_0 \sim 20~\mathrm{cm}^{-3}$.

In this scenario, particles accelerated at the WTS first propagate within the bubble, then diffuse through the cloud, and finally escape into the interstellar medium (ISM). The diffusion coefficient in the bubble-cloud system can then be written as 
\begin{equation}
    D_2(r,E) = 
    \begin{cases}
    D_{\mathrm{b}}(r,E), & R_{\mathrm{s}}<r \leq R_{\mathrm{b}} \\
    D_{\mathrm{c}}(E), & R_{\mathrm{b}}<r \leq R_{\mathrm{c}},
    \end{cases}
\end{equation}
where $D_{\mathrm{b}}$ and $D_{\mathrm{c}}$ are the diffusion coefficients in the bubble and in the cloud, respectively. We take $D_{\mathrm{b}}$ to be the non-uniform Bohm diffusion coefficient introduced above. For $D_{\mathrm{c}}$, we start by adopting the typical value of the diffusion coefficient in the ISM, given empirically by $4\times 10^{28} ~ E_{\mathrm{GeV}}^{1/3} ~ \mathrm{cm}^2 ~ \mathrm{s}^{-1}$ \citep{Evoli_2019,Gabici_2019}, and assume it to be uniform throughout the cloud with a radius of $R_{\mathrm{c}} \sim$ 150 pc. 

We expect that high-energy GCRs may penetrate the bubble diffusively, while lower-energy particles, whose propagation is dominated by advection, are unable to reach the inner regions of the bubble. For our parameter values, the boundary between the two regimes is in the region of $\sim$ 10 TeV. In order to compute the effect of GCR penetration in the bubble, we solved the transport equation imposing the boundary conditions that there is not net flux at $r=0$ and that $N(R_{\mathrm{c}},E)=N_{\mathrm{GCR}}(E)$, meaning that the density at the edge of the cloud equals the density of GCRs. We also assumed that the distribution of GCRs is uniform throughout the Galaxy and adopted the locally measured CR proton spectrum as the CR spectrum in our calculations. While the proton CR spectrum is known with good accuracy at energies of $\lesssim$10 TeV, the data at higher energies suffer from considerable systematics. Here we adopted the IceTop and LHAASO data \citep{Aartsen_2019,LHAASO_2025} and used a fit to the whole spectrum. The result of the fit is shown in Figure \ref{fig:gcr} (see \citep{LHAASO_2024,Harer_2025} for other estimates of the contribution of GCRs to the gamma-ray emission of the Cygnus bubble).

Before continuing with this discussion, we want to stress that some CRs might be residing inside the bubble and cloud from the time of formation of the SC, although any conclusion on this subject is bound to be model-dependent: if a collective wind is formed and the reduced diffusivity is due to conversion of a fraction of the wind ram pressure to magnetic energy, then we can expect that within one advection time (at most) all particles are ejected from the bubble, either due to advection or diffusion. At high energies, where diffusion dominates, there will be an equal flux of particles leaving and entering the bubble, and the density of CRs in the bubble is, to a good approximation, the same as we estimate by describing what we refer to as the CR penetration from the ISM. As for the cloud, where no advection is expected, particles can only evolve by diffusion. If the diffusivity is suppressed in the cloud, as we deduce below, then at energies below $\sim 10$ TeV the timescale for pp energy losses in a gas density of $\sim 20~\rm cm^{-3}$ is shorter than the diffusion time for escaping the cloud. As a result, pre-existing CRs will be depleted by energy losses for energies of $\lesssim$10 TeV, while higher-energy cosmic rays will be unaffected and have the same density as outside.

Figure \ref{fig:fast & slow diffusion}(a) shows the predicted gamma-ray flux considering both the shock-accelerated particles and the GCRs penetrating the bubble (including re-acceleration), in the case in which the diffusion coefficient in the bubble is the Galactic one. In these conditions virtually all of the GCRs penetrate the cloud and enter the bubble. However, the advection of the shocked wind inhibits the penetration of GCRs with an energy of $\lesssim$10 TeV upwind of the TS. This implies that re-acceleration of GCRs at the WTS is limited to particles with an energy of $\gtrsim$10 TeV, and has therefore a rather small effect. GCRs penetrating the cloud have enough interactions with the local gas to cause an appreciable gamma-ray emission in the energy range of $\gtrsim$300 TeV. On the other hand, since their spectrum is rather steep, the gamma-ray emission in the $\sim$ GeV range due to interactions in the cloud is large enough that the Fermi data are exceeded. However, we think it is worth mentioning that a comparison of our model, including GCRs, with Fermi data, is not completely straightforward, because of the analysis procedure adopted by the Fermi collaboration, which includes fitting and subtracting the contribution of GCRs from extended sources \citep{Abdollahi_2020}.

Figure \ref{fig:fast & slow diffusion}(f) shows the expected gamma-ray emission in the case in which the diffusivity in the cloud is suppressed ($D_{\mathrm{c}}(E) = 4 \times 10^{24} ~ E_{\mathrm{GeV}}^{0.6} ~ \mathrm{cm}^{2} ~ \mathrm{s}^{-1}$): this assumption implies that only particles with large enough energy can penetrate the whole cloud, while most particles in the GCRs only penetrate a layer of thickness $\sim \sqrt{2D_\mathrm{c}(E)T_\mathrm{age}}$, where $T_\mathrm{age} \simeq 3$ Myr is the age of the cluster. Here the densities in the bubble and in the cloud are chosen to be $n_\mathrm{b}=10~\rm cm^{-3}$ and $n_\mathrm{c}=20~\rm cm^{-3}$.

The expected morphology of the gamma-ray emission for the case of Galactic diffusion in the cloud, in the energy range accessible to HAWC ($>1$ TeV), is shown in Figure \ref{fig:fast & slow diffusion}(b) together with the corresponding morphology to be compared with LHAASO data at higher energies (Figure \ref{fig:fast & slow diffusion}(c-e)). The same plots for the case of suppressed diffusion in the cloud are shown in Figure \ref{fig:fast & slow diffusion}(g-j), where a better agreement with data can be inferred. 

Notice that for the morphology to be properly accounted for it is important for the gas density inside the bubble to be comparable to that of the ambient medium. This can be seen in Figure \ref{fig:low density bubble}, where we show the results for a small diffusion coefficient in the cloud but for a bubble density\footnote{In the absence of thermal conduction, the gas density within the bubble is expected to be comparable to the shocked wind density, $n_\mathrm{w} \simeq 3\dot{M} T_\mathrm{age}/(4\pi R_\mathrm{b}^3 m_\mathrm{H}) \sim 0.005 \,\mathrm{cm}^{-3}$, which is significantly smaller than unity. However, for illustrative purposes, we adopt $n_\mathrm{b}=1~\mathrm{cm}^{-3}$.} of $n_\mathrm{b}=1~\rm cm^{-3}$: while the spectrum is equally well reproduced, the resulting morphology is quite unlike the observed one, as a result of the more rarefied bubble. This is an important point, in that situations in which most of the CR interactions occur in the outer region are likely to reproduce the spectrum well but to fail to explain the morphology of the gamma-ray emission.
It is worth noting that the above discussion assumes that the bubble is surrounded by a cloud with a density of $20 \, \mathrm{cm}^{-3}$, which is consistent with gas observations, albeit with large uncertainties \citep{Astiasarain_2023,Menchiari_2023,Menchiari_2024,Harer_2025}. In Appendix \ref{app:no cloud}, we also consider a bubble evolving in a more dilute ISM, with a density of $\sim 2 \, \mathrm{cm}^{-3}$. We show that it remains true that a good description of the data requires that the density inside the bubble be within a factor of $\sim 2$ of the density outside.

We want to stress again that our best model for the LHAASO gamma-ray spectrum and morphology, as shown in Figure \ref{fig:fast & slow diffusion}(f-j), required several extreme assumptions. For instance, in the calculations above, we assumed $\eta_B=0.1$, which corresponds to an Alfv\'enic shock Mach number of $M_{\mathrm{A}} = \eta_B^{-1/2} \approx 3$. This might be problematic for particle acceleration. Hence it is legitimate to wonder how the results would change for lower values of $\eta_B$. Figure \ref{fig:lower etaB} shows our results for $\eta_B=0.02$. It is clear that a lower $\eta_B$ reduces the maximum energy that can be reached at the WTS. In order to account for the gamma-ray flux in the energy range of $\gtrsim$100 TeV, one could enhance the contribution from GCRs by adopting a higher gas density in the cloud of $n_\mathrm{c}=40~\rm cm^{-3}$, while keeping $n_\mathrm{b}=10~\rm cm^{-3}$. Although it is still possible to fit the gamma-ray flux, the morphology at energies of >100 TeV cannot be reproduced. We conclude that a relatively large value of $\eta_B$ is needed to explain both the gamma-ray flux and morphology of the Cygnus bubble. This adds to the severe requirements of a model based on particle acceleration at the collective wind.

\section{Point source with constant injection in the centre of the star cluster}
\label{sec:cont}

Here we consider the case of a hypothetical constant source in the centre of the SC, injecting CRs into the surrounding region. Such a source has to be active as an accelerator for times at least comparable with the age of the cluster itself. In principle, the winds of individual stars inside the cluster or the interaction regions between two or more such winds might be sites of acceleration \citep{Voelk_1982,Cesarsky_1983,Webb_1985,Parizot_2004,Aharonian_2019}, although, as argued by \cite{Harer_2025}, it is unlikely that these accelerators can energize particles to $\gtrsim$100 TeV energies, based on the simple Hillas criterion. By assumption, we apply this scenario to a situation in which the collective wind of the SC is not formed, as proposed by \cite{Vieu_2024}. Here we speculate that one or more such regions accelerate particles up to $E_\mathrm{max}\sim$1 PeV, and that these particles diffuse in the region around the SC. The injection spectrum is assumed to be in the form $N_{\mathrm{inj}}(E) \propto E^{-s} ~ \mathrm{exp}(-E/E_{\mathrm{max}})$. 

Before performing detailed calculations, it is useful to provide some preliminary estimates of the requirements on the diffusion coefficient $D(E)=D_0 E_{\mathrm{GeV}}^{\delta}$ that we need to impose in order to reconcile predictions and gamma-ray data. The low-energy gamma-ray spectrum, in the Fermi-LAT region, shows a relatively hard photon index, $\Gamma_1 \sim 2.1$, while the HAWC and LHAASO measurements in the TeV energy range suggest a gradual steepening with an approximate slope of $\Gamma_2 \sim 2.64$ and 2.6--2.93, respectively \citep{Abdollahi_2020,Abeysekara_2021,LHAASO_2024}. In the context of a purely diffusive model, this transition is naturally accounted for in terms of particles that during the age of the cluster are confined in the region of interest (at low energies), versus the ones (at higher energies) that cannot be confined. By imposing that the transition between the two regimes occurs around $\sim$ 10 TeV for the primary CR protons, one obtains the following estimate for the diffusion coefficient: 
\begin{equation}
D(E=10~\mathrm{TeV})\approx 2.3\times 10^{27}~\mathrm{cm}^2~\mathrm{s}^{-1} ~\left(\frac{R}{150~\mathrm{pc}}\right)^2~\left(\frac{T_{\mathrm{age}}}{3~\mathrm{Myr}}\right)^{-1}.
\nonumber
\end{equation}
The diffusion spectral index can be approximated as $\delta \sim \Delta \Gamma=\Gamma_2 - \Gamma_1 \sim$ 0.5--0.8.

The case of a continuous source in the centre of the cluster, with the inclusion of pp energy losses, was treated by numerically solving the transport equation in its time dependent version, using a Crank-Nicholson scheme. Since there is no TS to be accounted for, this calculation is rather standard. The integration region that we considered is made of a bubble around the cluster and a surrounding cloud of material outside the bubble. At the edge of the cloud, the boundary condition is that the spectrum of CRs is the GCR spectrum. This boundary condition allows us to properly describe the penetration of GCRs into the system. Estimating the size of the bubble is non-trivial in a case in which there is a hot outflow with no WTS. Here we assume that even in the absence of a supersonic outflow we can use the value of $R_\mathrm{b}$ in Appendix \ref{app:evolution} as an estimate of the size, keeping in mind that in the end the bubble is set by the pressure balance between inside and outside. The need for a cloud of material outside the bubble is imposed by the gamma-ray morphology extending to $\sim 6$ degrees, appreciably larger than the size of the bubble. For simplicity we assumed that the diffusion coefficient in the bubble and the cloud is the same. The penetration of GCRs into the bubble and of the accelerated particles into the cloud is properly described by following the temporal evolution of the particle spectrum over the age of the SC. 

\begin{figure}[t]
    \centering
    \begin{subfigure}[b]{0.48\textwidth} 
    \centering
    \includegraphics[width=\textwidth]{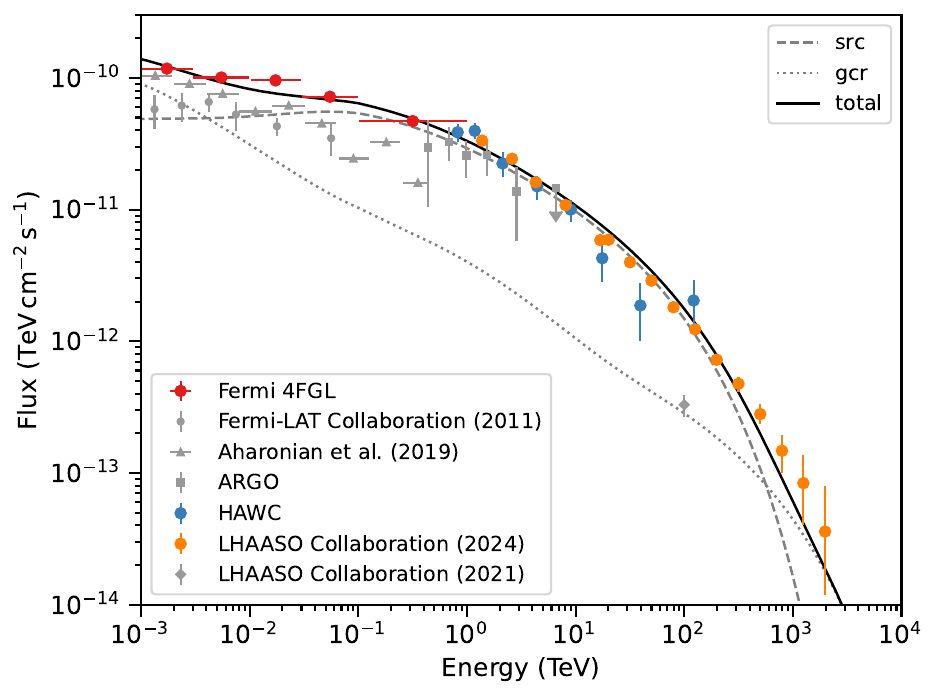} 
    \end{subfigure}
    
    \begin{subfigure}[b]{0.48\textwidth}
    \centering
    \includegraphics[width=\textwidth]{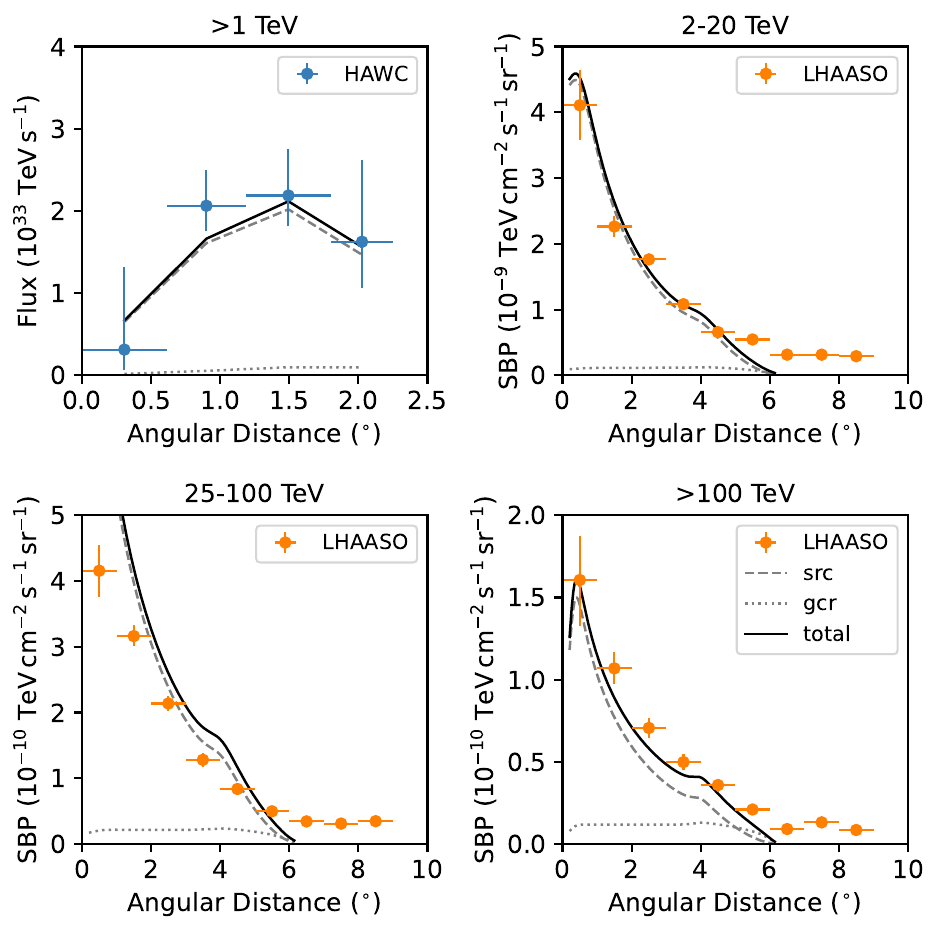} 
    \end{subfigure}
    \caption{Continuous injection from a central point source, assuming a bubble gas density of $n_{\rm b}=10 ~ \rm{cm}^{-3}$ and a diffusion coefficient with $D_0=4\times 10^{24}~\rm cm^2 ~ s^{-1}$ and $\delta=0.6$.}
    \label{fig:cont1}
\end{figure}

In Figure \ref{fig:cont1} we show the results of our calculations for the spectrum (top panel) and morphology (bottom panels) in the case of a point source constantly releasing particles over the course of the cluster's existence. In the top panel, the dashed and dotted lines represent, respectively, the gamma-ray emission due to CRs accelerated in the putative source and the gamma rays produced by GCRs penetrating the region. The density in the bubble and the cloud are chosen as $n_\mathrm{b}=10~\rm cm^{-3}$ and $n_\mathrm{c}=20~\rm cm^{-3}$, respectively. The injection spectrum required to fit the data has a slope of $s=2$, a maximum energy of $E_\mathrm{max}=$1 PeV, and a luminosity of $L_{\mathrm{CR}}=2.6\times 10^{36}~\mathrm{erg}~\mathrm{s}^{-1}$ in the form of accelerated protons. The adopted diffusion coefficient is $D_0=4 \times 10^{24} ~ \mathrm{cm}^2 ~ \mathrm{s}^{-1}$ and $\delta=0.6$. The solid line indicates the total gamma-ray emission. One can appreciate that despite the very high maximum energy of locally accelerated particles, the GCR interactions are needed to account for the highest-energy data points. The morphology in all the energy bands is satisfactorily accounted for with this set of parameters.  

While we did not specify the nature of the potential accelerators, the necessary luminosity and adopted value of $E_{\rm max}$ impose some non-trivial constraints. Indeed, the maximum energy achievable in a source of luminosity $L_{\rm source}=L_{\rm CR}/\xi_{\rm CR}$ and characterized by a flow speed, $v$, can be written as (Hillas criterion)
\begin{align}
E_{\rm max} &= e \sqrt{\frac{\eta_B v L_{\rm source}}{c^2}} = e \sqrt{\frac{\eta_B v L_{\rm CR}}{\xi_{\rm CR} c^2}} \\
& \simeq 100\ {\rm TeV}\ \left(\frac{\eta_B}{\xi_{\rm CR}}\right)^{1/2} \left(\frac{v}{10^3 ~ {\rm km ~ s^{-1}}}\right)^{1/2} \left(\frac{L_{\rm CR}}{10^{36} ~ {\rm erg ~ s^{-1}}}\right)^{1/2}, \nonumber
\end{align}
where $\eta_B<1$ is the fraction of the source kinetic luminosity that is converted into magnetic field energy. It is then clear that, in order for a source with $L_{\rm CR}=10^{36} \, \mathrm{erg} \, \mathrm{s}^{-1}$ to accelerate CRs to PeV energy the condition
\begin{equation}
\left(\frac{\eta_B}{\xi_{\rm CR}}\right) \left(\frac{v}{10^3 \, \mathrm{km} \, \mathrm{s}^{-1}}\right) \simeq 100
\label{eq:howhard}
\end{equation}
must be satisfied. This implies that in the case of a Wolf-Rayet wind, with kinetic luminosity $\lesssim 10^{37} \, \mathrm{erg} \, \mathrm{s}^{-1}$ and flow velocity $2\times 10^{3} \, \mathrm{km} \, \mathrm{s}^{-1}$ at most, acceleration to PeV energy would need $\eta_B\gtrsim 5$. This is equivalent to stating that it is impossible for this class of sources to be the origin of PeV radiation.

If we assume that $\eta_B/\xi_{\mathrm{CR}}\lesssim 1$, which is usually the case, then Equation \eqref{eq:howhard} suggests that in order to reach PeV energies the velocity must exceed $\sim 10^5\, \mathrm{km} \, \mathrm{s}^{-1}$, hinting at relativistic sources. One may then wonder whether the Cygnus association might host an unidentified pulsar wind nebula or microquasar.

\begin{figure}[t]
    \centering
    \begin{subfigure}[b]{0.48\textwidth} 
    \centering
    \includegraphics[width=\textwidth]{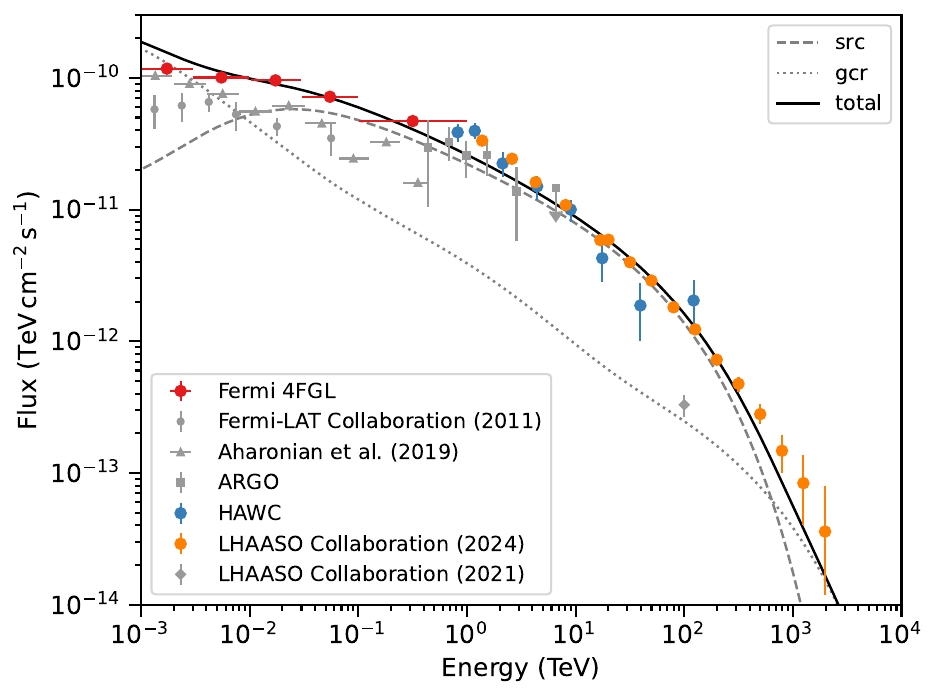} 
    \end{subfigure}
    
    \begin{subfigure}[b]{0.48\textwidth}
    \centering
    \includegraphics[width=\textwidth]{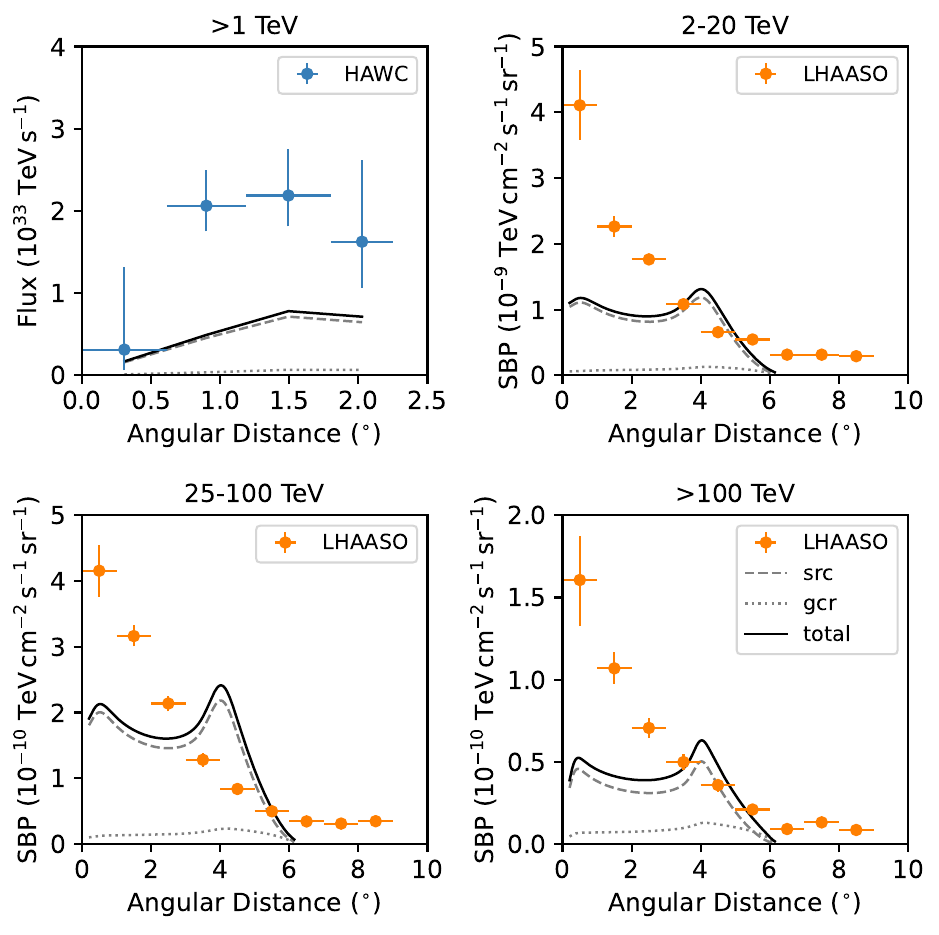} 
    \end{subfigure}
    \caption{Same as Figure \ref{fig:cont1}, but with a lower bubble gas density of $n_{\rm b}=1 ~ \rm{cm}^{-3}$ and a diffusion coefficient with $D_0=2\times 10^{25}~\rm cm^2 ~ s^{-1}$ and $\delta=0.5$.}
    \label{fig:cont2}
\end{figure}

In order to explain the importance of a relatively high gas density (in the form of dense clouds) inside the bubble, in Figure \ref{fig:cont2} we show the results of our calculations for the case of a mean gas density of $n_\mathrm{b}=1~\rm cm^{-3}$. The fitted injection luminosity and diffusion coefficient become $6\times 10^{36}~\mathrm{erg}~\mathrm{s}^{-1}$ and $D_0=2 \times 10^{25} ~ \mathrm{cm}^2 ~ \mathrm{s}^{-1}$ with $\delta=0.5$. While it is still possible to provide a satisfactory fit to the total gamma-ray spectrum, the morphology that arises from this scenario is quite unlike the data, as a clear consequence of depleting the gamma-ray emission from the central part of the system. In this case, as well as in the case of particle acceleration at the WTS, the morphology of the gamma-ray emission can only be explained by adopting a relatively high gas density in the bubble. Moreover in both cases a suppressed diffusion coefficient in both the bubble and the cloud is required in order to account for the gamma-ray spectrum. 

\section{Point source with burst-like injection in the centre of the star cluster}
\label{sec:burst}

\begin{figure}[t]
    \centering
    \begin{subfigure}[b]{0.48\textwidth} 
    \centering
    \includegraphics[width=\textwidth]{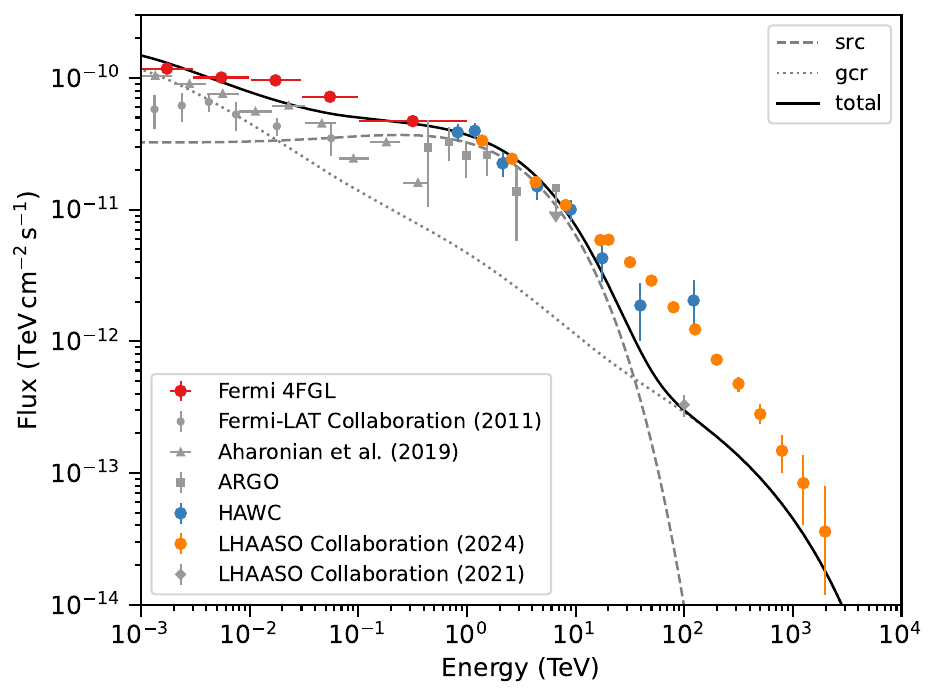} 
    \end{subfigure}
    
    \begin{subfigure}[b]{0.48\textwidth}
    \centering
    \includegraphics[width=\textwidth]{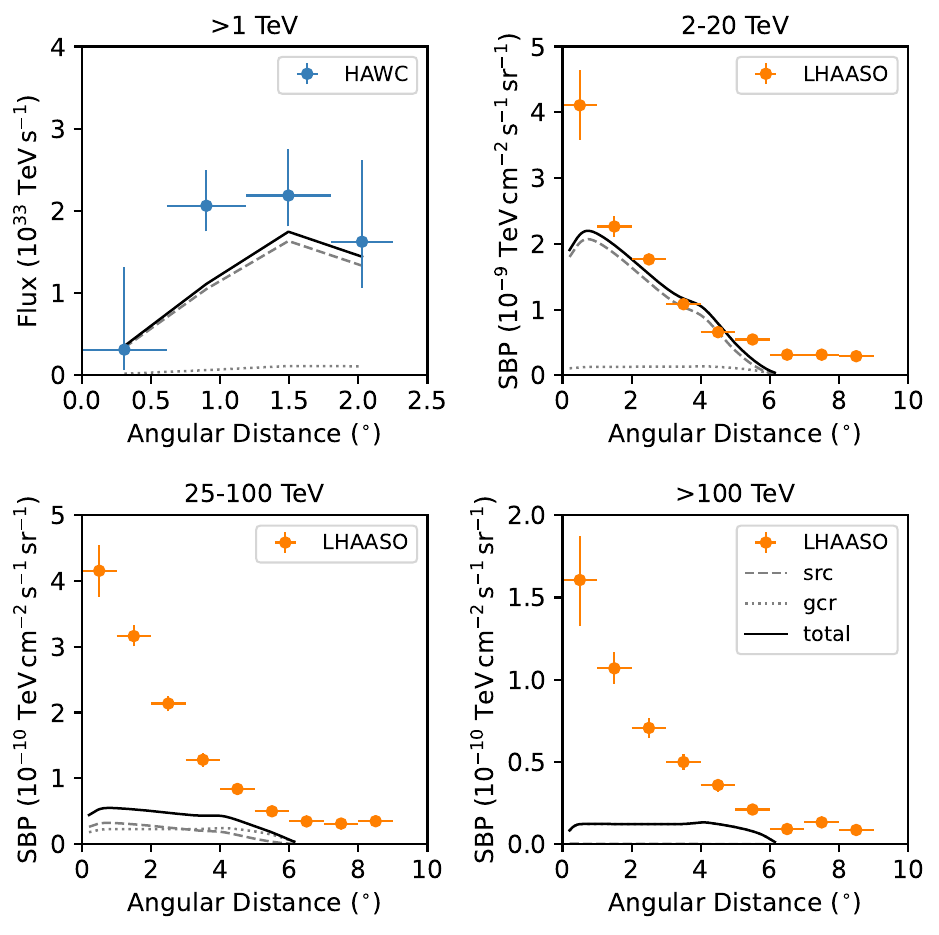} 
    \end{subfigure}
    \caption{Impulsive injection from a central point source, assuming a bubble gas density of $n_{\rm b}=10 ~ \rm{cm}^{-3}$ and a diffusion coefficient with $D_0=4\times 10^{24}~\rm cm^2 ~ s^{-1}$ and $\delta=0.8$.}
    \label{fig:burst1}
\end{figure}

\begin{figure}[t]
    \centering
    \begin{subfigure}[b]{0.48\textwidth} 
    \centering
    \includegraphics[width=\textwidth]{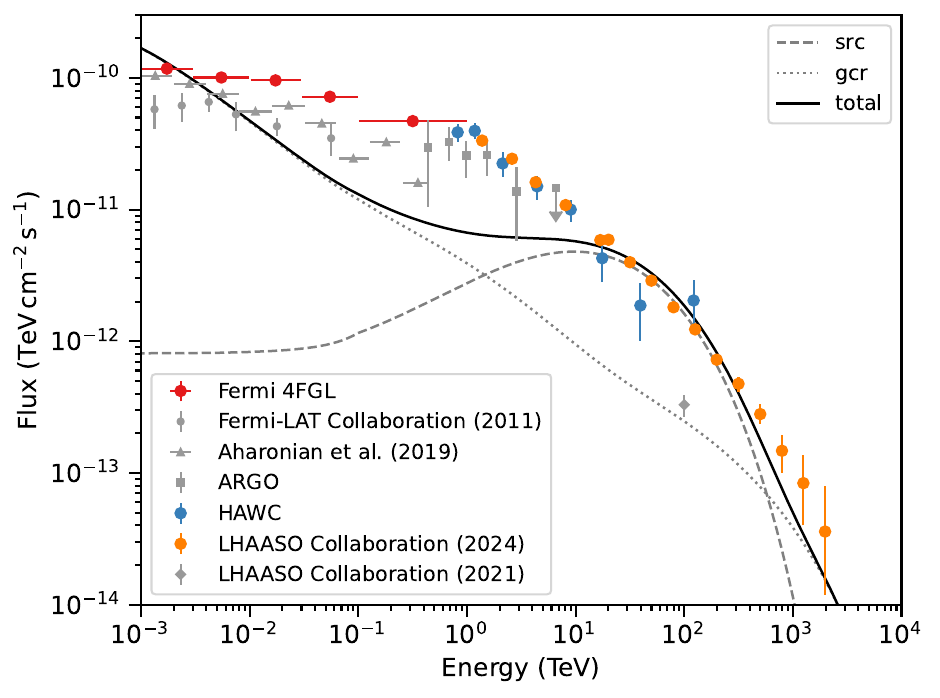} 
    \end{subfigure}
    
    \begin{subfigure}[b]{0.48\textwidth}
    \centering
    \includegraphics[width=\textwidth]{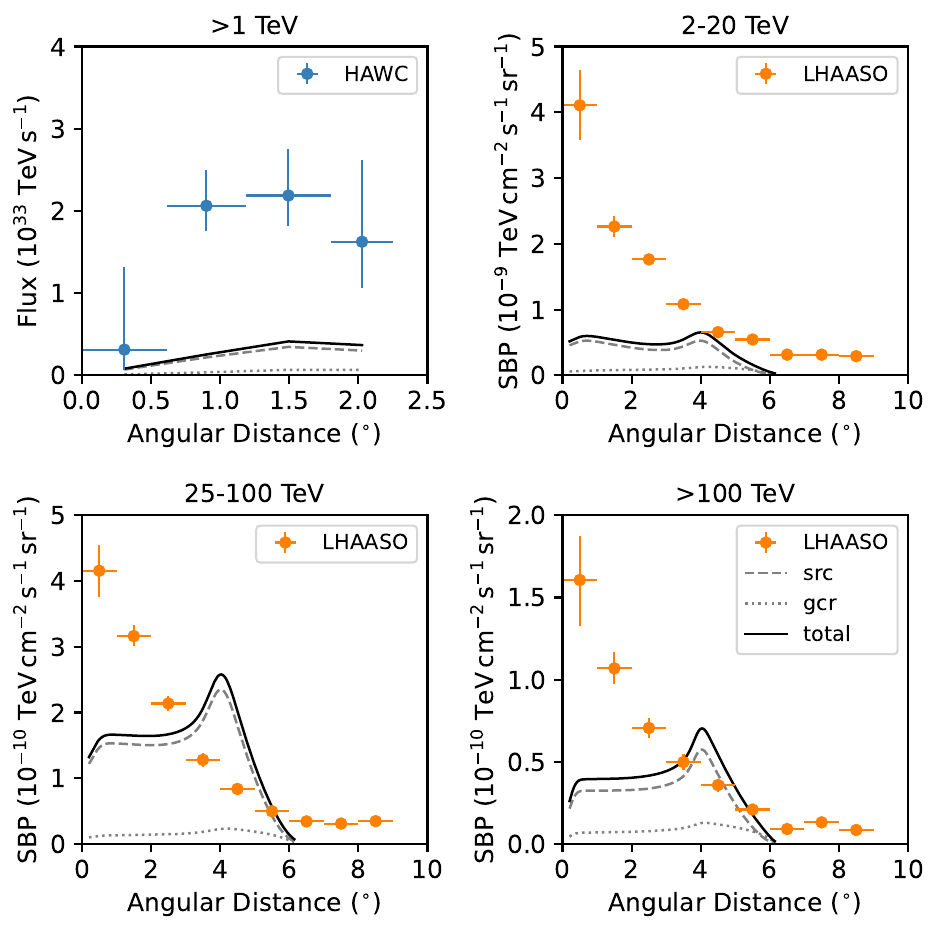} 
    \end{subfigure}
    \caption{Same as Figure \ref{fig:burst1}, but with a lower bubble gas density of $n_{\rm b}=1 ~ \rm{cm}^{-3}$ and a diffusion coefficient with $D_0=2\times 10^{25}~\rm cm^2 ~ s^{-1}$ and $\delta=0.5$.}
    \label{fig:burst2}
\end{figure}

The age of the Cygnus SC adopted here, three million years, is somewhat questionable based on the fact that the cluster seems to contain substructures of different ages and that one such structure might be old enough to have hosted a supernova explosion. In fact the presence of a pulsar wind nebula in the Cygnus field can be used to support this possibility. Such a hypothetical supernova cannot have occurred too recently, otherwise we would likely observe its remnant, and it cannot have happened too long ago, otherwise the accelerated particles would have escaped the SC. Here we speculate on the possibility that a supernova event may have occurred a few tens of thousands of years ago, and describe the gamma-ray emission that is produced by CR protons accelerated at the forward shock of the remnant. We stress here that, as discussed by \cite{Sushch2025}, the maximum energy at such a supernova remnant (SNR) is unlikely to reach the PeV region, unless effective magnetic field amplification occurs and acceleration takes place in the Bohm regime. These are basically the same limitations that appear to limit $E_\mathrm{max}$ to values much smaller than PeV energy in an ordinary SNR. The case of a supernova explosion was also recently discussed by \cite{Harer_2025}, who estimated the maximum energy by using the standard Hillas criterion. For the purposes of this calculation, we assume that the supernova event is burst-like and solve the time-dependent transport equation with the standard boundary conditions of zero net flux at $r=0$ and $N(E,r=R_\mathrm{c})=N_\mathrm{GCR}(E)$ at the edge of the cloud. As in the case of a stationary point-like source, these boundary conditions allow us to also take into account the penetration of GCRs into the system. 

In Figure \ref{fig:burst1} we show the gamma-ray emission and morphology for a supernova exploding 60 kyr ago, with a total energy of $10^{51}$ erg, a spectrum $\sim E^{-2}$ and $E_\mathrm{max}=$1 PeV, and an acceleration efficiency $\xi_\mathrm{CR}=0.1$. The gas densities in the bubble and in the cloud are $n_\mathrm{b}=10~\mathrm{cm}^{-3}$ and $n_\mathrm{c}=20~\mathrm{cm}^{-3}$. As in previous cases, the diffusion coefficient in the bubble is required to be substantially suppressed ($D_0=4\times 10^{24}~\rm cm^2 ~ s^{-1}$ and $\delta=0.8$) in order to account for any appreciable gamma-ray emission.  
This situation corresponds to an escape time of CRs from the region that is shorter than the age of the remnant at energies above $\sim$ 10 TeV, so that higher-energy particles have already left the system. The gamma-ray emission due to CRs accelerated at the SNR is therefore limited to energies $\lesssim$10 TeV. One should appreciate how the contribution to the gamma-ray emission due to the interactions of GCRs with gas in the cloud is hard to avoid: whether the diffusion coefficient in the cloud is suppressed or not, sufficiently high-energy GCRs would be able to penetrate the whole cloud and their density in the cloud would be basically the same as in the ISM. For the parameters adopted here, the emission of the cloud above 10 TeV is basically independent of the diffusion coefficient. At low energies, in the GeV range, in the absence of suppressed diffusion, the contribution of GCRs would be even larger than is shown in Figure \ref{fig:burst1}, because GCRs would penetrate through the whole cloud instead of a fraction of it.

It is clear from the top panel of Figure \ref{fig:burst1} that the gamma-ray spectrum cannot be accounted for in this scenario, hinting at the fact that other sources or other mechanisms need to be invoked. A similar conclusion was found by \cite{Harer_2025}, who suggested that the observed gamma-ray emission may have both a hadronic and a leptonic component, with IC emission from accelerated electrons providing a substantial contribution below tens of TeV. In such a scenario, electrons would need to be accelerated with an efficiency that is much larger than that usually associated with fast shocks (a typical ratio of $\sim 10^{-2}$ between accelerated electrons and protons \cite[]{DamianoElec2015}, as also required based on Galactic transport of protons and electrons and observations at the Earth \cite[]{Armillotta25,EvoliElec2021}).
As is shown in the bottom panels of Figure \ref{fig:burst1}, not only the spectrum but also the morphology of the gamma-ray emission is poorly reproduced, especially at high energies.

For completeness, we also computed the gamma-ray spectrum in the case of a lower density in the bubble, $n_\mathrm{b}=1~\rm cm^{-3}$, and the results are shown in Figure \ref{fig:burst2} for $\xi_\mathrm{CR}=0.025$ and $D_0=2\times 10^{25}~\rm cm^2 ~ s^{-1}$ with $\delta=0.5$. As expected, the flux of gamma radiation from the bubble drops, but in this setup the high-energy particles manage to escape into the cloud where gamma-ray production occurs. This results in the bump at $E_\gamma\gtrsim$100 GeV visible in Figure \ref{fig:burst2}. It remains true that the fit to both the spectrum and morphology is far from acceptable, implying the need for additional sources or processes for gamma-ray emission \citep{Harer_2025}.
We stress again that the contribution to gamma rays due to the interactions of GCRs with the cloud cannot be avoided and should be taken into account in any attempt to model these systems.

\section{Discussion and conclusions}
\label{sec:discuss}

The detection of very high-energy gamma rays from the Cygnus region, extending up to PeV energies, has triggered much interest in the topic of particle acceleration and transport in young SCs, especially in relation to the possibility that they may work as PeVatrons. We emphasize that in all cases discussed here, one or more extreme assumptions must be made in order to accommodate the LHAASO data, making the case of Cygnus very challenging. Here we discussed three scenarios for the acceleration and transport in the Cygnus region, aimed at providing a viable explanation of the spectrum and morphology of the gamma-ray emission as measured by Fermi-LAT, HAWC and LHAASO. The first scenario is based on the formation of a collective wind plowing material outward and excavating a rarefied hot bubble around the SC, with a size of $\sim 100$ pc for typical parameters of the Cygnus association. Diffusive particle acceleration takes place at the WTS \cite[]{Morlino_2021,Blasi_2023,Menchiari_2023,Menchiari_2024} and the accelerated particles are transported downstream through advection and diffusion. The spherical symmetry of the problem, for reasonable choices of the diffusion coefficient in the bubble, leads to a gamma-ray spectrum that drops too quickly with energy and is unable to account for the LHAASO data. In fact we show that the only case for which such data can be reasonably well described is that of space-dependent Bohm diffusion, where the strength of the local magnetic field downstream is computed as a fraction of the local ram pressure. We consider the assumption of Bohm diffusion to be highly speculative and unlikely to be realized in this context, in that it requires magnetic power to be independent of scale. This assumption may be realized when turbulence is self-generated by a spectrum of accelerated particles, $\propto E^{-2}$, but as discussed by \cite{Blasi_2023}, in Cygnus there do not seem to be favourable conditions for this process to take place. 

In the case of spatially dependent Bohm diffusion, a hadronic interpretation of the LHAASO data on morphology requires the gas density inside the bubble to be relatively large and not much smaller than the ambient medium, which might be in the form of the molecular gas from which the SC originally formed. In terms of modelling the gamma-ray emission from Cygnus, the presence of the cloud helps to explain the emission profiles extending out to $\sim 6$ degrees and, perhaps equally important, it allows us to speculate that interactions of GCRs in the cloud may account for gamma-ray emission with $E_\gamma \gtrsim 300~\mathrm{TeV}$.

The diffusion coefficients in both the bubble and the cloud are required to be suppressed with respect to their value in the Galaxy by several orders of magnitude in order to account for observations. This last finding is common to all the scenarios investigated here. 

Although the formation of a WTS is common wisdom in the theory of young SCs, it has been argued \cite[]{Vieu_2024} that the Cygnus cluster might be too loose and have too few young stars to form a collective WTS, so that smaller TSs, associated with the winds of individual stars, might be more likely to occur. For this reason, as well as due to the severe requirements of the WTS model described above, we also consider the case of a point source in the centre of the SC, be it continuously operating in time or rather burst-like. The continuous case might be associated with particle acceleration in the winds of individual stars in the cluster or the interaction between these winds. It should be noted that both these possibilities appear to require rather extreme and possibly unrealistic conditions for CR acceleration to PeV energies, as has also been discussed by \cite{Harer_2025}. This scenario also requires a bubble density of order $\sim 10~\rm cm^{-3}$, and the presence of an extended cloud region reaching out to $\sim$ 150 pc, in order to accommodate the observed morphology. As in the case of the WTS, here too the diffusion coefficient must be suppressed by orders of magnitude to achieve a good description of the spectrum. The contribution of the GCR interactions in the surroundings could play an important role both at very high energies due to the cut-off of the locally accelerated particle spectrum, and in the GeV range, as a result of the steep spectrum of GCRs. 

The last scenario investigated here is that of a burst-like source in the centre, such as could be the case of a supernova explosion. The event cannot have occurred too recently because we would then see the remnant, and it cannot have occurred too far in the past, since in that case all high-energy particles would have left the region, even in the case of suppressed diffusion. We tried to fit the data with a supernova event that occurred $\sim$ 50--60 kyr ago and for different choices of the diffusion coefficient in the bubble. In none of the cases is it possible to achieve a decent description of the data at all energies, in terms of either the spectrum or the morphology, confirming the previous results of \cite{Harer_2025}, despite the requirement that the diffusion coefficient in the region be suppressed (otherwise unreasonably large efficiencies are required) and despite the assumption of a dense extended cloud. We also stress the fact that, contrary to the claims of \cite{Harer_2025}, reaching $\gtrsim$ PeV energies in a supernova in Cygnus is equally difficult as in other situations: while Hillas criterion sets an upper limit to $E_\mathrm{max}$, the expected level of magnetic field amplification required in order to accelerate to such energies is still lacking theoretical (and observational) support (see also discussion in \cite{Sushch2025}).

The acceleration of leptons might lead to gamma-ray emission through inverse Compton scattering of background photons, as has been argued by \cite{Harer_2023} in the case of Westerlund 1. Typically, however, these models require a substantial fraction of the accelerator energy channeled into the electron component, quite unlike what we see in the case of particle acceleration in SNRs.  

The summary of our findings discussed above leads to a few conclusions. First, the spectrum and morphology of the gamma-ray emission of Cygnus as measured by LHAASO pose serious problems in terms of the acceleration of these particles. Moreover, in all scenarios discussed above there are a few common conclusions to be drawn: the diffusion coefficient in the bubble needs to be suppressed with respect to Galactic values by several orders of magnitude. The physical processes responsible for such an effect are currently unknown and would deserve proper attention even as a stand-alone problem. 

The second conclusion we can draw is that in order to accommodate the extended morphology observed by LHAASO, in all scenarios above the models require gas with a mean density of $\sim 10~\mathrm{cm}^{-3}$ extending out to $\sim 150$ pc from the cluster (see, however, Appendix \ref{app:no cloud}). While this may be molecular gas from which the SC formed a few million years ago, it remains to be understood what the effect is of the GCR interactions on such gas. The implications of such interactions in terms of discrimination between truly diffuse Galactic gamma-ray emission and near-source gamma-ray emission are huge, not to mention the role of such gas in terms of the near-source grammage felt by GCRs (see for instance \cite{Blasi2025,Cocoon2025}) and reflecting on secondary-to-primary ratios.

Finally, we want to comment on the fact that the relatively poor angular resolution of the current gamma-ray telescopes does not allow us to infer whether at least some of the observed gamma-ray emission from the direction of Cygnus may be due to unresolved sources either inside the cluster or behind it. In principle, there are two sources in the region that belong to classes of well-known powerful accelerators: the pulsar J2032+4127 (and its wind nebula), and the $\mu$QSO Cygnus X-3. However, PSR J2032+4127 has been excluded both on theoretical and observational grounds: soon after the Cygnus region was first detected by LHAASO, the pulsar was shown to not be powerful enough to provide the needed potential drop for PeV particle acceleration \cite[]{deOna2022}; later on, a detailed LHAASO analysis of the region \cite[]{LHAASO_2024} showed that the spectrum of this source has a cut-off at a few tens of TeV. In addition, the $\mu$QSO Cygnus X-3 is a prominent gamma-ray source whose emission could potentially be extended and provide a contribution to the diffuse emission from the Cygnus region. Recently, LHAASO has detected variable gamma-ray emission up to $\sim$ 3.7 PeV from Cygnus X-3 \citep{LHAASO_cygx3}. Using a duty cycle of 0.4 as suggested by LHAASO data, and using the estimated average gamma-ray flux in the PeV region from this source (estimated to be $\sim 2 \times 10^{-14} ~ \mathrm{TeV} ~ \mathrm{cm}^2 ~ \mathrm{s}^{-1}$ at 1 PeV), we conclude that it is possible that $20-30\%$ of the gamma-ray emission from the Cygnus association can be attributed to Cygnus X-3 instead.
Moreover, this source is observed as point-like, probably due to the limited angular resolution of LHAASO. Future observations with a better angular resolution, for example with ASTRI Mini-Array \cite[]{ASTRI2022} and CTAO \cite[]{CTA2013}, will provide precious information and help us identify the truly diffuse emission from the Cygnus cocoon.

\begin{acknowledgements}
This work has been partially funded by the European Union - Next Generation EU under the project IR0000012---CTA+ ``Rafforzamento e creazione di IR nell’ambito del Piano Nazionale di Ripresa e Resilienza (PNRR)'', under PRIN-MUR 2022TJW4EJ ``Unveiling the footprints of the cosmic ray journey through the Galaxy and beyond'' and under MUR National Innovation Ecosystem grant ECS00000041 - VITALITY/ASTRA - CUP D13C21000430001. 
\end{acknowledgements}

\bibliographystyle{aa}
\bibliography{bib_ben}

\begin{appendix}

\section{Superbubble evolution}
\label{app:evolution}

The interaction between the collective wind of a compact SC and its ambient medium excavates an interstellar bubble. The gas swept by the forward shock is expected to form a thin shell around the bubble, with a total mass $M_{\mathrm{sh}} = 4\pi R_{\mathrm{b}}^3(t) \rho_0/3$, where $\rho_0$ is the mass density of the ambient medium and is assumed to be uniform here. The motion of the shell is governed by Newton's law, namely
\begin{equation}
    \frac{d}{dt} \left[ M_{\mathrm{sh}}(t) \dot{R}_{\mathrm{b}}(t) \right] = 4\pi R_{\mathrm{b}}^2(t) P,
    \label{eq:momentum}
\end{equation}
where $P$ is the shocked wind pressure.

The SC continuously injects kinetic energy into the bubble. If we neglect the thermal conduction and radiative cooling of the hot gas in the bubble, according to the law of energy conservation, the total energy in the bubble evolves as 
\begin{equation}
    \frac{d}{dt} \left[ \frac{4}{3} \pi R_{\mathrm{b}}^3(t) \frac{P}{\gamma - 1} \right] = L_{\mathrm{w}} - 4 \pi R_{\mathrm{b}}^2(t) P \dot{R}_{\mathrm{b}}(t),
    \label{eq:energy}
\end{equation}
where $L_{\mathrm{w}}=(1/2)\dot{M} v_{\mathrm{w}}^2$ is the wind kinetic luminosity, with $\dot{M}$ and $v_{\mathrm{w}}$ the mass loss rate of the SC and the wind velocity. $P/(\gamma-1)$ is the internal energy density of the hot bubble, where the adiabatic index is $\gamma=5/3$ for an ideal monatomic gas.

Equations \eqref{eq:momentum}--\eqref{eq:energy} can be solved by assuming $R_{\mathrm{b}} \propto t^{\alpha}$, which gives \citep{Dyson_1972, Castor_1975, Weaver_1977, Gupta_2018,Morlino_2021}
\begin{equation}
    R_{\mathrm{b}}(t) \simeq 0.76 ~ L_{\mathrm{w}}^{1/5} \rho_{0}^{-1/5} t^{3/5}, \ \mathrm{and}
\end{equation}
\begin{equation}
    P(t) \simeq 0.16 ~ L_{\mathrm{w}}^{2/5} \rho_{0}^{3/5} t^{-4/5}.
\end{equation}
On the other hand, the location of the TS can be calculated by balancing the pressure in the bubble and the ram pressure of the wind, which yields
\begin{equation}
    R_{\mathrm{s}}(t) = \left(\frac{\dot{M}v_{\mathrm{w}}}{4\pi P}\right)^{1/2} \simeq 0.7 \dot{M}^{1/2} v_{\mathrm{w}}^{1/2} L_{\mathrm{w}}^{-1/5} \rho_{0}^{-3/10} t^{2/5}.
\end{equation}

\section{Particle transport in the superbubble}
\label{app:transport}

Non-thermal particles propagate by diffusion and advection in the bubble excavated by the cluster wind. Their energy and spatial distributions can be obtained by solving the transport equation. Assuming spherical symmetry of the bubble, the transport equation is expressed as
\begin{equation}
    \frac{1}{r^2} \frac{\partial}{\partial r} \left[r^2 D \frac{\partial f}{\partial r}\right] - \frac{1}{r^2} \frac{\partial}{\partial r} \left[r^2\tilde{u} f\right] + \frac{1}{p^2} \frac{\partial}{\partial p} \left[p^2\dot{p} f\right] + Q = \frac{\partial f}{\partial t},
\end{equation}
where $f(r,p,t)$ is the particle distribution function in phase space. $Q(r,p,t)$ is the injection term, which can be expressed by $Q(r,p,t)=4 \pi R_{\mathrm{s}}^2 \eta_{\mathrm{inj}} n_1 u_1 [\delta(p - p_{\mathrm{inj}})/(4 \pi p^2)] [\delta(r - R_{\mathrm{s}})/(4 \pi r^2)]$. The energy loss term includes both the adiabatic energy loss/gain and the energy loss by interactions, namely $\dot{p}=\dot{p}_{\mathrm{ad}}+\dot{p}_{\mathrm{int}}$. $D(r,p,t)$ and $\tilde{u}(r,t)$ are the diffusion coefficient and the advection velocity, respectively.

Assuming the system has reached a steady state, the transport equation for relativistic protons can be written as
\begin{equation}
    \frac{1}{r^2} \frac{\partial}{\partial r} \left[r^2 D \frac{\partial N}{\partial r}\right] - \frac{1}{r^2} \frac{\partial}{\partial r} \left[r^2\tilde{u} N\right] + \frac{\partial}{\partial E} \left[\dot{E} N\right] + Q = 0,
\label{eq:transport2}
\end{equation}
where $N(r,E)$ is the distribution function in energy space, such that $N(r,E) dE=f(r,p) 4 \pi p^2 dp$. The injection term becomes $Q(r,E)=4 \pi R_{\mathrm{s}}^2 \eta_{\mathrm{inj}} n_1 u_1 \delta(E - E_{\mathrm{inj}}) [\delta(r - R_{\mathrm{s}})/(4 \pi r^2)]$, with $E_{\mathrm{inj}}$ the injection energy.
The diffusion coefficient can be written as 
\begin{equation}
    D(r,E) = 
    \begin{cases}
    D_1(r,E), & 0<r \leq R_{\mathrm{s}} \\
    D_2(r,E), & R_{\mathrm{s}}<r \leq R_{\mathrm{b}}
    \end{cases}
\end{equation}
where $D_1$ and $D_2$ are the diffusion coefficients upstream and downstream of the shock respectively. For a detailed discussion of the diffusion coefficients, see Section \ref{sec:max energy}.
The advection velocity of non-thermal particles can be expressed as 
\begin{equation}
    \tilde u(r) = 
    \begin{cases}
    \tilde u_1, & 0<r \leq R_{\mathrm{s}} \\
    \tilde u_2 \left(\frac{r}{R_{\mathrm{s}}}\right)^{-2}, & R_{\mathrm{s}}<r \leq R_{\mathrm{b}}
    \end{cases}
\end{equation}
where $\tilde u_1$ and $\tilde u_2$ are the effective velocities immediately upstream and downstream of the shock. The effective velocity is defined as the sum of the plasma speed $u(r)$ and the net speed of the turbulent waves that act as scattering centres, $\tilde u(r)=u(r)+\eta v_{\mathrm{A}}$, where $v_{\mathrm{A}}=B/\sqrt{4\pi \rho}$ is the Alfv\'en speed. Here we assume that waves propagate isotropically in the upstream ($\eta_1=0$), while anisotropy can exist in the downstream ($\eta_2\neq0$). 
The energy loss rate can be written as the sum of the adiabatic energy loss rate and the energy loss rate of pp interactions, $\dot{E}=\dot{E}_{\mathrm{ad}}+\dot{E}_{\mathrm{int}}$, with
\begin{equation}
    \dot{E}_{\mathrm{ad}}=\frac{1}{3} \left( \nabla \cdot \bm{\tilde u} \right) E, \ \mathrm{and}
\end{equation}
\begin{equation}
\begin{aligned}
    \dot{E}_{\mathrm{int}} \simeq \
    & 5.1 \times 10^{-15} K_{\pi} n_\mathrm{b} \left( \frac{E}{\mathrm{GeV}} \right) \\
    & \times \left[ 1 + 5.5 \times 10^{-2} L + 7.3 \times 10^{-3} L^2 \right] ~ \mathrm{GeV} ~ \mathrm{s}^{-1},
\end{aligned}
\end{equation}
where $L=\ln \left(E/\mathrm{TeV}\right)$ and $n_\mathrm{b}$ is the number density of target gas in the bubble. $K_{\pi} \simeq 0.13$ is adopted as a good approximation \citep{Krakau_2015}.
Boundary conditions are required to solve Equation \eqref{eq:transport2}. We assume that the particle flux is zero at the inner boundary, namely
\begin{equation}
    \left[u N-D \frac{\partial N}{\partial r}\right]_{R_{\mathrm{sc}}}=0, 
\end{equation}
with $R_{\mathrm{sc}}$ the size of the SC,
and that particles can freely escape the system from the outer boundary: $N(R_{\mathrm{b}},E)=0$.

To obtain the particle spectrum at the shock, we first substitute the adiabatic energy loss rate into Equation \eqref{eq:transport2}, so that the transport equation becomes
\begin{equation}
\begin{aligned}
    \frac{\partial}{\partial r} \left[r^2 \left(\tilde{u}N - D \frac{\partial N}{\partial r}\right) \right] =
    &\frac{1}{3} \frac{d(r^2\tilde{u})}{dr} \frac{\partial}{\partial E}\left(EN\right) + r^2 \frac{\partial}{\partial E}\left(\dot{E}_{\mathrm{int}}N\right) \\
    &+ R_{\mathrm{s}}^2 \eta_{\mathrm{inj}} n_1 u_1 \delta(E - E_{\mathrm{inj}}) \delta(r - R_{\mathrm{s}}).
\end{aligned}
\label{eq:transport3}
\end{equation}
We then integrate Equation \eqref{eq:transport3} from $R_{\mathrm{s}}^{-}$ to $R_{\mathrm{s}}^{+}$, which yields
\begin{equation}
\begin{aligned}
    -D_2 \frac{\partial N}{\partial r} \bigg|_2 + D_1 \frac{\partial N}{\partial r} \bigg|_1 = 
    &\frac{1}{3} (\tilde{u}_2 - \tilde{u}_1) E \frac{dN_0}{dE} - \frac{2}{3} (\tilde{u}_2 - \tilde{u}_1) N_0 \\ 
    &+ \eta_{\mathrm{inj}} n_1 u_1 \delta(E - E_{\mathrm{inj}}).
\end{aligned}
\label{eq:transport4}
\end{equation}
By introducing
\begin{equation}
    D_{1,2} \frac{\partial N}{\partial r} \bigg|_{1,2} = \mathcal{D}_{1,2}(E) \tilde{u}_{1,2} N_0(E),    
\end{equation}
Equation \eqref{eq:transport4} becomes
\begin{equation}
\begin{aligned}
    &\frac{dN_0}{dE} + \left[ \frac{3\mathcal{D}_2(E)}{1-\mathcal{R}} - \frac{3\mathcal{R}(\mathcal{D}_1(E) - 1)}{1-\mathcal{R}} + \frac{\mathcal{R} + 2}{\mathcal{R} - 1} \right] \frac{N_0}{E} \\
    &= \frac{3u_1}{\tilde u_1-\tilde u_2} \frac{\eta_{\mathrm{inj}} n_1 \delta(E - E_{\mathrm{inj}})}{E},
\end{aligned}
\label{eq:transport5}
\end{equation}
where $\mathcal{R}=\tilde u_1 / \tilde u_2$ is the effective compression factor. The solution for $N_0(E)$ at $E>E_{\mathrm{inj}}$ can be expressed as 
\begin{equation}
   N_0(E) = K \left(\frac{E}{E_{\mathrm{inj}}}\right)^{-\frac{\mathcal{R}+2}{\mathcal{R}-1}} \times \exp \left\{ \int_{E_{\mathrm{inj}}}^{E} \frac{dE'}{E'} \left[\frac{3\mathcal{R}(\mathcal{D}_{1}-1)}{1-\mathcal{R}} - \frac{3\mathcal{D}_2}{1-\mathcal{R}}\right] \right\},
\label{eq:solution}
\end{equation}
with
\begin{equation}
    K= \frac{3u_1}{\tilde u_1-\tilde u_2} \frac{\eta_{\mathrm{inj}} n_1}{E_{\mathrm{inj}}}.
\end{equation}
The spectrum at the shock is normalized by an efficiency of conversion ($\xi_{\mathrm{CR}}$) of the wind ram pressure to the pressure of non-thermal particles.
Equation \eqref{eq:solution} is an implicit solution for the particle spectrum at the shock, since $\mathcal{D}_{1,2}(E)$ depends on the diffusive fluxes immediately upstream and downstream of the shock, which are unknown a priori. To solve for $N(r,E)$, we adopt the numerical iterative algorithm developed by \cite{Blasi_2023}. 

\section{Gamma-ray emission from the superbubble}
\label{app:emission}

The gamma-ray emissivity from pp interactions can be calculated as (in units of $\mathrm{cm}^{-3} ~ \mathrm{s}^{-1} ~ \mathrm{TeV}^{-1}$)
\begin{equation}
    J_{\gamma}(r, E_{\gamma}) = cn_\mathrm{b} \int_{E_\gamma}^{\infty} \sigma_{\mathrm{inel}}(E) N(r,E) F_\gamma \left( \frac{E_\gamma}{E}, E \right) \frac{dE}{E}, 
\end{equation}
where $F_{\gamma}$ is the gamma-ray spectrum per collision, and $\sigma_{\mathrm{inel}}$ is the inelastic cross section of pp interactions \citep{Kelner_2006,Kafexhiu_2014}. 

The gamma-ray intensity is then obtained by integrating the gamma-ray emissivity along the line of sight
\begin{equation}
    I_{\gamma}(\theta, E_{\gamma}) = \frac{1}{4\pi} \int_{l_{\mathrm{-}}}^{l_{\mathrm{+}}} J_{\gamma}\left[r(\theta,l), E_{\gamma}\right] dl,
\end{equation}
where $\theta$ is the angle between the line of sight and the direction toward the centre of the bubble, and $l$ is the distance from the observer to a point along the line of sight, with $l_{\mathrm{\pm}}=d \cos{\theta} \pm \sqrt{R_{\mathrm{b}}^2-d^2 \sin^2{\theta}}$. The integration is divided by $4\pi$ to account for the assumption of isotropic emission. To compare with the gamma-ray morphologies measured by LHAASO, the predicted gamma-ray intensity is convolved with the point spread function (PSF) of the instrument \citep{Zhang_2021,Bao_2022,Hu_2024}.
The total gamma-ray flux from the bubble is finally calculated as
\begin{equation}
    \Phi_{\gamma}(E_{\gamma}) = 2\pi \int I_{\gamma}(\theta, E_{\gamma}) \sin{\theta} d\theta,
\end{equation}
in units of $\mathrm{cm}^{-2} ~ \mathrm{s}^{-1} ~ \mathrm{TeV}^{-1}$.

\section{The case without a surrounding cloud}
\label{app:no cloud}

Here we present the results in the case where there is no cloud surrounding the Cygnus bubble. In order to match the bubble size with the gamma-ray observations (i.e, $\sim 150~\mathrm{pc}$), the ambient gas density is required to be $\sim 2~\mathrm{cm}^{-3}$, as can be easily deduced from Equation \eqref{eq:bubble radius}. This density is comparable to that of the ISM and therefore represents a lower limit for the surrounding gas density (Observations of the gas distribution indicate a gas density of order $\sim 10~\mathrm{cm}^{-3}$ \citep{Astiasarain_2023,Menchiari_2023,Menchiari_2024,Harer_2025}). In this case, the cluster wind interacts directly with the ISM. We note that, if radiative cooling of the bubble is taken into account, its size would be $\sim 70\%$ smaller, thus requiring an even lower ambient gas density to reproduce the observed size.

\begin{figure}[t]
    \centering
    \begin{subfigure}[b]{0.48\textwidth} 
    \centering
    \includegraphics[width=\textwidth]{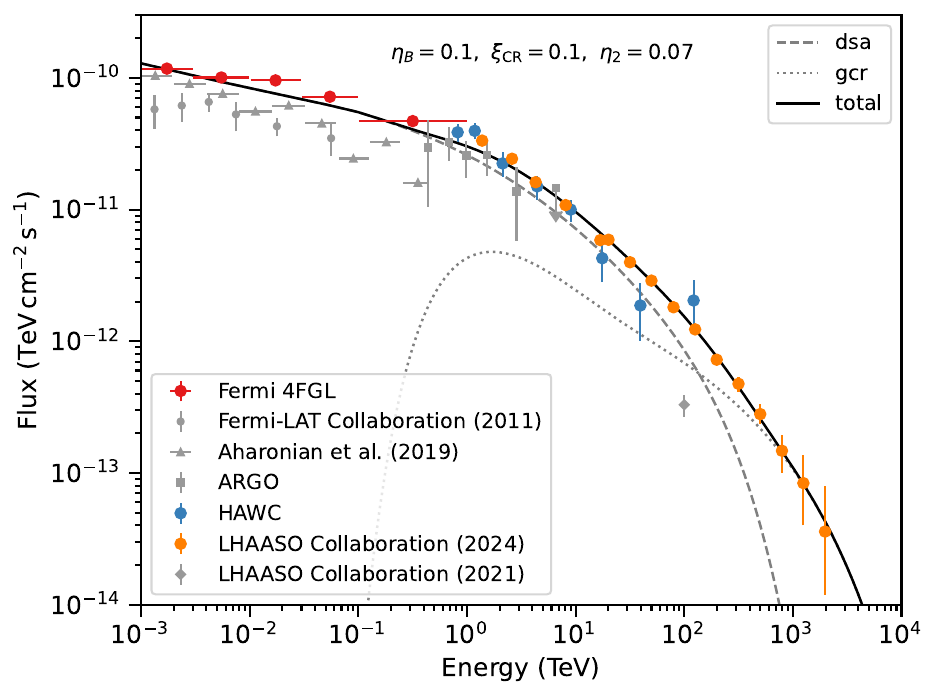} 
    \end{subfigure}
    
    \begin{subfigure}[b]{0.48\textwidth}
    \centering
    \includegraphics[width=\textwidth]{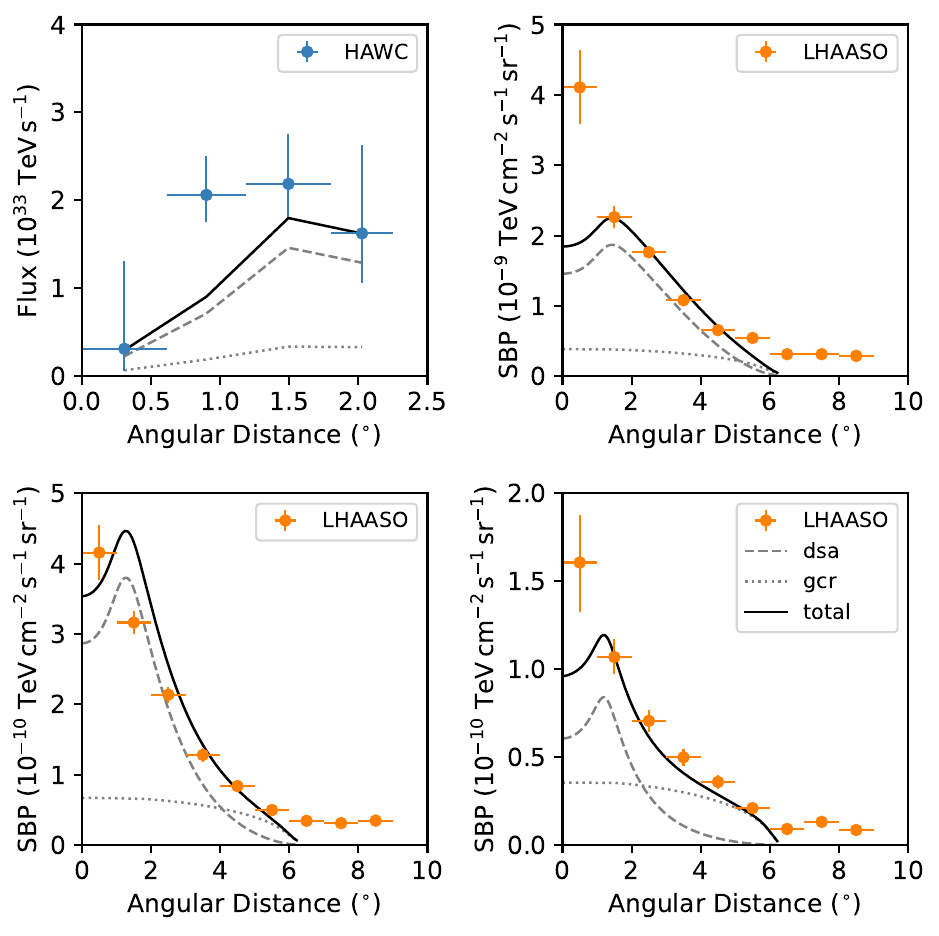} 
    \end{subfigure}
    \caption{Particle acceleration at the WTS, assuming a bubble density of $n_{\rm b}=1 ~ \rm{cm}^{-3}$ and an ambient density of $n_0=2 ~ \rm{cm}^{-3}$ (resulting in $R_\mathrm{b} \sim 150~\mathrm{pc}$). A spatially dependent Bohm diffusion is adopted. The diffuse gamma-ray emission from GCRs is calculated to reproduce the observed PeV gamma-ray flux by adjusting the gas column density along the line of sight. A low-energy cutoff at $\sim 1~\mathrm{TeV}$ is applied to the diffuse gamma-ray flux to mimic the subtraction of the diffuse component in the GeV data analysis of Fermi-LAT.}
    \label{fig:no cloud}
\end{figure}

Figure \ref{fig:no cloud} shows the results in the absence of a cloud. Here, we adopt a gas density of $n_\mathrm{b}=1~\mathrm{cm}^{-3}$ inside the bubble and $n_0=2~\mathrm{cm}^{-3}$ outside. A spatially dependent Bohm diffusion is assumed. It is clear that the gamma-ray emission from protons accelerated at the WTS can hardly reach PeV energies. On the other hand, the gamma-ray flux produced by GCRs penetrating the bubble is negligible due to the small gas density in this case. In order to accommodate the observed PeV gamma-ray flux, we calculate the total diffuse gamma-ray emission from GCRs by adjusting the gas column density along the line of sight in the direction of the Cygnus bubble. We also apply an artificial low-energy cutoff at $\sim 1~\mathrm{TeV}$ to the diffuse gamma-ray flux to emulate the subtraction of the diffuse component in the GeV data analysis of Fermi-LAT \citep{Abdollahi_2020}. 

After considering the diffuse gamma-ray emission, both the gamma-ray spectrum and morphology can be nicely reproduced except for data points in the innermost $\sim 1$ degree region, implying an additional unresolved point source in this region (e.g., Cygnus X-3 \citep{LHAASO_cygx3}). The adopted acceleration efficiency is $\xi_\mathrm{CR}=0.1$, which indicates that the bubble gas density should be at least $n_\mathrm{b}\sim 1~\mathrm{cm}^{-3}$ in order to account for the observed gamma-ray flux. This density is comparable to the ambient gas density $n_0\sim 2~\mathrm{cm}^{-3}$, and is appreciably higher than the shocked wind density, i.e., $n_\mathrm{w} \simeq 3\dot{M} T_\mathrm{age}/(4\pi R_\mathrm{b}^3 m_\mathrm{H}) \sim 10^{-3} \, \mathrm{cm}^{-3}$, indicating significant mass loading from the dense, cool shell due to thermal conduction, or survival of clumpy clouds after the passage of the wind forward shock. We note that the bubble gas density adopted here should be regarded as a limiting case. The actual gas density in this region may be of order $\sim 10~\mathrm{cm}^{-3}$. A higher ambient gas density will reduce the bubble size, again implying the presence of a surrounding cloud (as in the case considered in the main text). Moreover, suppressed particle diffusion over a region extending to at least $\sim 150~\mathrm{pc}$, together with a relatively high magnetic efficiency $\eta_B$, is still required to account for both the observed spectrum and morphology.

\end{appendix}

\end{document}